\documentstyle[epsf]{mn}

\newif\ifAMStwofonts

\AMStwofontstrue

\title[RR Lyrae variables in the SMC - II]{RR Lyrae variables in the Small Magellanic Cloud - II. The extended area: chemical and structural analysis}
\author[E.~Kapakos and D.~Hatzidimitriou]
       {E.~Kapakos$^1$\footnotemark and D.~Hatzidimitriou$^{1}$\\
$^1$Department of Physics - Section of Astrophysics, Astronomy \& Mechanics, University of Athens, 157 84 Zografos, Athens, Greece\\}

\date{Accepted 2012 July 30 ;  Received 2012 July 28 ; in original form 2012 March 6}

\pagerange{\pageref{firstpage}--\pageref{lastpage}}

\pubyear{2012}

\begin{document}

\label{firstpage}

\maketitle

\begin{abstract}

\noindent We have performed the Fourier decomposition analysis of 8- and 13-year V-band light curves of a carefully selected sample of 454 fundamental-mode RR Lyrae variables (RRab type), detected in a $\simeq 14$ square degree area of the Small Magellanic Cloud (SMC) and listed in the Optical Gravitational Lensing Experiment, phase III, Catalogue of Variable Stars. The Fourier decomposition parameters were used to derive metal abundances and distance moduli, following the methodology described by Kapakos, Hatzidimitriou \& Soszy\'{n}ski. The average metal abundance of the RRab stars on the new scale of Carretta et al. was found to be $\langle [Fe/H]_{C09} \rangle=-1.69\pm0.41$ dex (std, with a standard error of $0.02$ dex). A tentative metallicity gradient of $-0.013\pm0.007$ dex/kpc was detected, with increasing metal abundance towards the dynamical center of the SMC, but selection effects are also discussed. The distance modulus of the SMC was re-estimated and was found to be $\langle \mu \rangle=19.13\pm0.19$ (std) in a distance scale where the distance modulus of the Large Magellanic Cloud (LMC) is $\mu_{LMC} = 18.52\pm0.06$(std). The average $1\sigma$ line-of-sight depth was found to be $\sigma_{int}=5.3\pm0.4$ kpc (std), while spatial variations of the depth were detected. The SMC was found to be deeper in the north-eastern region, while metal richer and metal poorer objects in the sample seem to belong to different dynamical structures. The former have smaller scale height and may constitute a thick disk, its width being $10.40\pm0.02$ kpc, and a bulge whose size (radius) is estimated to be $2.09\pm0.81$ kpc. The latter seem to belong to a halo structure with a maximum depth along the line of sight extending over 16 kpc in the SMC central region and falling to $\sim 12$ kpc in the outer regions.

\end{abstract}

\begin{keywords}

methods: data analysis - astronomical databases: miscellaneous - stars: abundances - stars: variables: RR Lyrae - Magellanic Clouds

\end{keywords}

\footnotetext[1]{E-mail: efkap@phys.uoa.gr}
\renewcommand{\thefootnote}{\arabic{footnote}}


\section{Introduction}

This is the second paper in a series presenting photometric metal abundances of RR Lyrae variable stars in the Magellanic Clouds and their use for chemical and structural analysis. In Kapakos, Hatzidimitriou \& Soszy\'{n}ski (2011, hereinafter Paper I), the results of the Fourier decomposition analysis of the V-band light curves of 84 fundamental-mode and 16 first overtone pulsators in the central bar region of the Small Magellanic Cloud (SMC) were presented. The methodology was described in detail with full discussion of the method, limitations, biases and sources of error. However, the spatial extent of the sample in the central bar region did not allow the investigation of important issues such as a possible metallicity gradient, which has been proposed by other independent studies (e.g. Carrera et al. 2009). Furthermore, studies on the morphology of the SMC seem to reveal interesting features. The structure of the SMC is affected by interactions with both the LMC and our Galaxy (e.g. Hatzidimitriou \& Hawkins 1989, hereinafter HH89; Bekki 2009). Subramanian \& Subramaniam (2009, hereinafter SS09) have found its depth along the line-of-sight (LOS) to be larger than that of the LMC and a bulge-like structure near the optical center, based on red horizontal branch (red clump) stars. A larger depth in the outer region has also been suggested, probably partially connected to a tidal stream-like structure in the north-eastern outer regions (Hatzidimitriou, Cannon \& Hawkins 1993). Although the distribution of old red stars in the SMC (i.e. giants and red clump stars with ages $\geq1 Gyr$, according to Zaritsky et al. 2000) and the velocity field of red giant branch stars, which does not show much rotation (Harris \& Zaritsky 2006), suggest a spheroidal structure, HI and young stars are found to have ordered rotation and to reside in a disk (van der Marel, Kallivayalil \& Besla 2009). In Paper I, we found indications that old populations as represented by RR Lyrae stars are distributed in two dynamical structures with different average metal abundances. Thus, a study of a larger sample distributed over a larger area is needed to clarify all these issues.

\par{We have used the RR Lyrae variables detected in the extended region of the SMC with the Optical Gravitational Lensing Experiment (OGLE), phase III (OGLE-III\footnotemark[1] \footnotetext[1]{The full extent of the OGLE-III data set was not available for Paper I.}), combined with the data collected from phase II (OGLE-II) in the central region. Our aim is to derive metal abundances ([Fe/H]) and distance moduli for all RR Lyrae variables of RRab type with well observed light curves in order to proceed to a chemical and structural analysis of the old populations of this galaxy, using a much larger sample of objects distributed over a more extended area of the SMC than in Paper I. We decided to exclude RRc type stars from our analysis, since discrepancies associated with the calibrating equations of their Fourier parameters versus metallicity were detected in Paper I. However, we provide the results of the Fourier decomposition for these stars as well, for completeness reasons.}

\par{In Section 2, we describe the data used and give the Fourier decomposition results. In Section 3, we derive the metal abundances of a carefully selected sample of RRab stars and discuss their distribution, as well as the positions of these objects on the Bailey diagram. In Section 4, we investigate the existence of a metalicity gradient in the area covered by our study. In Section 5, we derive the absolute magnitudes and distance moduli of the RR Lyrae stars and we proceed to a structural analysis of the SMC. Variations to its LOS depth and structure-metallicity relations are examined. In Section 6, we summarize our results.}


\section{Data Analysis and Fourier Decomposition}

\subsection{Data description}

The present study is based on observations obtained with the 1.3m Warsaw telescope at the Las Campanas Observatory, Chile, during phase III of the OGLE project, between 2001 and 2009 (Soszy\'{n}ski et al. 2010, hereinafter Sos10). These observations contain precisely calibrated photometric and astrometric data in the I- and V-band filters, from the extended area of the SMC, covering $\simeq14$ square degrees on the sky in 41 fields of $\simeq35^{\prime}\times35^{\prime}.5$ (see Udalski 2003 and Udalski et al. 2008 for reduction pipeline, techniques and observational details). Although the data are thoroughly discussed in Sos10, some issues are worth mentioning. Stars located in the overlapping regions of adjacent fields were detected twice and their photometry was compiled from all available sources. Moreover, for RR Lyrae variables in the central bar region of the SMC, which was also covered by OGLE-II observations between 1997 and 2000, the two photometry databases were merged after the appropriate inter-calibrations, i.e. by shifting the OGLE-II photometry to match the OGLE-III light curves.

\par{Following Paper I, we opted for V-band photometry since the dependence of [Fe/H] on the Fourier parameters has been directly calibrated only for this band. It should be noted that some authors (e.g. Deb \& Singh 2010, hereinafter DS10; Haschke et al. 2012, hereinafter HGDJ) have used I-band data to derive metal abundances and physical parameters of RR Lyrae stars. As thoroughly discussed in Paper I, although the I-band light curves may produce similar average metallicities to those derived from the V-band light curves, discrepancies are revealed in metallicities of individual stars and their distributions. Furthermore, the errors of the Fourier parameters (and thus those of the corresponding metal abundances) are lower when using V-band light curves. The applicability of the calibrating equations is based on a deviation parameter introduced by Jurcsik \& Kov\'{a}cs (1996, hereinafter JK96; see Section 3). When the corresponding criterion is applied directly in V-band light curves, the selection effects are minimized.}

\par{There are 2475 RR Lyrae variables in the SMC, listed in the ninth part of the OGLE-III Catalogue of Variable Stars (OIII-CVS). The classification of the RR Lyraes as fundamental-mode pulsators (RRab), first overtone pulsators (RRc), double-mode pulsators (RRd) or second overtone pulsators (RRe), is described in Sos10. We have only retained RRab and RRc stars with V-band light curves, thus limiting our sample to 2092 stars, 1922 of them being of RRab type and 170 of RRc type. Among these stars, 23 objects (22 RRab and 1 RRc star) are Galactic foreground RR Lyrae variables and thus were excluded from further discussion. Another 13 RRab stars were discarded from the final data set, 12 having poorly populated time series (with less than $4m$ points, were m is the order of the fit) and 1 having high $\sigma_{fit}$ (see next subsection). The final sample consisted of 1887 RRab stars and 169 RRc stars, listed in Tables 1 and 2, respectively, along with properties of their light curves which are described later in this Section.}

\par{We have adopted the periods provided in OIII-CVS. The RRab and RRc variables in our sample have average periods of $0.60\pm0.06$ days (std) and $0.37\pm0.03$ days (std), respectively, the latter being expectedly shorter than the former. The individual periods and corresponding errors are listed in Column (2) of Tables 1 \& 2 for the RRab and RRc types respectively. It should be noted that these periods have been re-derived in the OGLE-III release and thus they are slightly different from the values used in Paper I for stars appearing both in OGLE-II and OGLE-III.}

\subsection{Fourier Decomposition}

\begin{table*}
\begin{minipage}{180mm}
\begin{center}\scriptsize
\caption{Fourier decomposition parameters for 1887 RRab stars derived from data of the OGLE phases II \& III in the V-band.}
\begin{tabular}{@{}ccccccccccc@{}}
\hline
OGLE Star ID&$P_0 (days)$&N&$\sigma_{fit}$&$A_0$&$A_1$&$A_3$&$A_V$&$\varphi_{31}$&$\varphi_{41}$&$D_m$\\
\hline
OGLE-SMC-RRLYR-0009&0.6947615(17)&78&0.053&19.477(3)&0.311(4)&0.114(4)&0.940&5.27(6) &1.82(9) &2.03(1.10)\\
OGLE-SMC-RRLYR-0029&0.5937182(9) &50&0.089&19.667(4)&0.269(6)&0.074(5)&0.704&4.75(44)&1.24(14)&3.92(1.39)\\
OGLE-SMC-RRLYR-0036&0.5222231(3) &51&0.089&19.646(4)&0.456(6)&0.133(6)&1.233&4.41(6) &0.80(8) &3.54(1.55)\\
OGLE-SMC-RRLYR-0048&0.5910502(7) &50&0.071&19.529(4)&0.289(5)&0.108(5)&0.832&4.58(8) &0.52(10)&2.05(1.63)\\
OGLE-SMC-RRLYR-0058&0.5704761(6) &51&0.072&19.816(5)&0.379(7)&0.088(7)&0.962&4.49(10)&0.77(15)&4.45(2.11)\\
OGLE-SMC-RRLYR-0059&0.4666863(8) &50&0.141&20.045(6)&0.378(9)&0.157(9)&1.228&4.48(9) &0.95(11)&4.96(2.15)\\
OGLE-SMC-RRLYR-0066&0.5676896(4) &50&0.062&19.357(3)&0.350(4)&0.144(4)&1.149&5.16(5) &1.69(7) &2.79(1.12)\\
OGLE-SMC-RRLYR-0075&0.5225272(4) &52&0.098&19.715(4)&0.425(6)&0.131(6)&1.151&4.48(7) &0.89(9) &3.56(1.49)\\
OGLE-SMC-RRLYR-0076&0.6641898(21)&47&0.063&19.697(5)&0.194(8)&0.063(7)&0.556&5.04(14)&1.30(96)&2.71(2.73)\\
OGLE-SMC-RRLYR-0077&0.6043705(12)&50&0.077&19.657(5)&0.274(7)&0.077(7)&0.708&4.77(11)&1.48(15)&3.85(2.83)\\
\\
Average value& &65&0.086&19.731(281)&0.285(89)&0.085(41)& &4.94(62)&1.47(95)& \\
\hline
\end{tabular}
\end{center}
\emph{Notes.} The numbers in parentheses are standard deviations multiplied by $10^6$ for the period, $10^3$ for $A_j$, $A_V$, 1 for $D_m$ and $10^2$ for the rest. The complete table is available in electronic form (see Supporting Information) and includes the original number of points $N_0$ in each light curve, all $A_j$ (where j=0,1,2,3,4), $R_{j1}$ and $\varphi_{j1}$ (where j=2,3,4).
\end{minipage}
\end{table*}

\begin{table*}
\begin{minipage}{180mm}
\begin{center}\scriptsize
\caption{Fourier decomposition parameters for 169 RRc stars derived from data of the OGLE phases II \& III in the V-band.}
\begin{tabular}{@{}cccccccccc@{}}
\hline
OGLE Star ID&$P_0 (days)$&N&$\sigma_{fit}$&$A_0$&$A_1$&$A_4$&$A_V$&$\varphi_{21}$&$\varphi_{31}$\\
\hline
OGLE-SMC-RRLYR-0020&0.3519121(7) &42&0.054&19.528(4)&0.245(5)&0.012(5)&0.494& 3.66( 23)&5.30(23)\\
OGLE-SMC-RRLYR-0035&0.3438972(10)&42&0.050&19.369(3)&0.257(4)&0.017(5)&0.532& 3.36( 15)&5.46(23)\\
OGLE-SMC-RRLYR-0040&0.3949114(8) &49&0.054&19.574(4)&0.261(6)&0.011(5)&0.545& 3.47( 16)&5.46(19)\\
OGLE-SMC-RRLYR-0050&0.3485816(5) &50&0.048&19.390(4)&0.269(5)&0.028(6)&0.593& 2.95( 12)&5.07(18)\\
OGLE-SMC-RRLYR-0062&0.3793270(8) &50&0.057&19.530(4)&0.239(4)&0.015(5)&0.530&-0.38(300)&6.48(19)\\
OGLE-SMC-RRLYR-0088&0.3444114(6) &51&0.074&19.732(5)&0.252(7)&0.054(7)&0.572& 3.02( 14)&7.02(14)\\
OGLE-SMC-RRLYR-0140&0.3694145(22)&50&0.105&19.656(4)&0.169(7)&0.039(7)&0.408& 0.93( 21)&2.82(29)\\
OGLE-SMC-RRLYR-0213&0.3975514(13)&48&0.083&19.663(4)&0.161(5)&0.015(5)&0.342& 2.05( 21)&6.74(13)\\
OGLE-SMC-RRLYR-0218&0.3532244(10)&47&0.055&19.625(5)&0.228(8)&0.013(6)&0.489& 2.79( 16)&5.43(28)\\
OGLE-SMC-RRLYR-0232&0.3383378(5) &71&0.051&19.419(2)&0.272(3)&0.007(3)&0.565& 2.88( 08)&4.98(22)\\
\\
Average value& &63&0.073&19.681(275)&0.236(43)&0.021(13)& &3.00(76)&5.56(146)\\
\hline
\end{tabular}
\end{center}
\emph{Notes.} The numbers in parentheses are standard deviations multiplied by $10^6$ for the period, $10^3$ for $A_j$, $A_V$ and $10^2$ for the rest. The complete table is available in electronic form (see Supporting Information) and includes the original number of points $N_0$ in each light curve, all $A_j$ (where j=0,1,2,3,4), $R_{j1}$ and $\varphi_{j1}$ (where j=2,3,4).
\end{minipage}
\end{table*}

The light curves of the 2056 RR Lyrae variables of RRab and RRc type in our final sample were fitted with fourth-order Fourier series of sine functions (equation 1 of Paper I or A1 in the Appendix\footnotemark[2] \footnotetext[2]{All equations of Paper I, which are used in the present work, are listed in the Appendix in order to facilitate the reading of this paper.}), following the technique adopted in Paper I, where there is a detailed discussion of the mathematical formula, the choice of the order of the fit and the criterion for excluding certain points from the light curves (which showed large deviations from the bulk of the data). All amplitudes $A_j$ and phases $\varphi_{j}$ (where $j=1,2,3,4$), as well as the ratios $R_{j1}=A_j / A_1$ and coefficients $\varphi_{j1}=\varphi_j-j\varphi_1$ (where $j=2,3,4$) were derived, while their standard deviations were estimated using Monte Carlo simulations and appropriate error propagation relations, as in Paper I.

\par{The results of the Fourier decomposition are presented in Tables 1 and 2 (the complete versions being available in electronic form) for RRab and RRc stars, respectively. In these Tables we also provide the number of points used for the fitting (N), the sigma of the fit for each light curve ($\sigma_{fit}$) and the amplitude in $V$ ($A_V$). The corresponding standard deviations of the various Fourier parameters are given in parentheses following each value. At the bottom of each table we give the average values of the relevant parameters for the entire sample with the associated standard deviations.}


\section{Metal Abundances}

\begin{table}
\begin{minipage}{85mm}
\begin{center}\scriptsize
\caption{Average values and the corresponding standard deviations of parameters derived using the Fourier decomposition technique for the final sample of 454 RR Lyrae variables of RRab type.}
\begin{tabular}{@{}cccc@{}}
\hline
Parameter                      & Value            & Parameter                      & Value         \\
\hline
$\langle \sigma_{fit} \rangle$ &  0.081$\pm$0.021 & $\langle R_{21} \rangle$       & 0.43$\pm$0.07 \\
                               &                  & $\langle R_{31} \rangle$       & 0.32$\pm$0.06 \\
$\langle A_0 \rangle$          & 19.72$\pm$0.20   & $\langle R_{41} \rangle$       & 0.20$\pm$0.06 \\
$\langle A_1 \rangle$          &  0.32$\pm$0.07   &                                &               \\
$\langle A_2 \rangle$          &  0.14$\pm$0.04   & $\langle \varphi_{21} \rangle$ & 2.32$\pm$0.17 \\
$\langle A_3 \rangle$          &  0.10$\pm$0.03   & $\langle \varphi_{31} \rangle$ & 4.92$\pm$0.30 \\
$\langle A_4 \rangle$          &  0.06$\pm$0.03   & $\langle \varphi_{41} \rangle$ & 1.40$\pm$0.44 \\
\\
$\langle A_V \rangle$          &  0.88$\pm$0.21   & $\langle D_m \rangle$          & 3.45$\pm$1.01 \\
\hline
\end{tabular}
\end{center}
\end{minipage}
\end{table}

\begin{table*}
\begin{minipage}{180mm}
\begin{center}\scriptsize
\caption{Metal abundances, absolute magnitudes, distance moduli and distances for the 454 RRab stars.}
\begin{tabular}{@{}cccccc@{}}
\hline
OGLE Star ID &${[Fe/H]}_{JK96} (dex)$&${[Fe/H]}_{C09} (dex)$&$M_V$&$\mu$      &$d$(kpc)     \\
\hline
OGLE-SMC-RRLYR-0009& -1.70(9)  & -1.82(14) & 0.47(7)  & 18.95(8)  & 61.58(2.16) \\
OGLE-SMC-RRLYR-0029& -1.85(59) & -1.96(60) & 0.44(14) & 19.17(14) & 68.20(4.54) \\
OGLE-SMC-RRLYR-0036& -1.92(9)  & -2.04(15) & 0.42(7)  & 19.16(8)  & 68.01(2.50) \\
OGLE-SMC-RRLYR-0048& -2.07(11) & -2.18(17) & 0.39(8)  & 19.08(8)  & 65.50(2.39) \\
OGLE-SMC-RRLYR-0058& -2.08(14) & -2.20(19) & 0.39(8)  & 19.38(8)  & 75.03(2.85) \\
OGLE-SMC-RRLYR-0059& -1.52(13) & -1.64(17) & 0.50(7)  & 19.48(7)  & 78.78(2.63) \\
OGLE-SMC-RRLYR-0066& -1.16(7)  & -1.27(12) & 0.58(7)  & 18.71(7)  & 55.24(1.86) \\
OGLE-SMC-RRLYR-0075& -1.70(10) & -1.81(15) & 0.47(7)  & 19.20(7)  & 69.29(2.34) \\
OGLE-SMC-RRLYR-0076& -1.85(19) & -1.96(23) & 0.44(8)  & 19.21(8)  & 69.43(2.66) \\
OGLE-SMC-RRLYR-0077& -1.88(15) & -2.00(20) & 0.43(8)  & 19.15(8)  & 67.64(2.54) \\
\hline
\end{tabular}
\end{center}
\emph{Notes.} The numbers in parentheses are standard deviations multiplied by $10^2$, except $\sigma_d$. The complete table is available in electronic form (see Supporting Information).
\end{minipage}
\end{table*}

\begin{table*}
\begin{minipage}{180mm}
\begin{center}\scriptsize
\caption{Average values for metal abundances, absolute magnitudes, distance moduli and distances for the 454 RRab stars.}
\begin{tabular}{@{}cccccc@{}}
\hline {\bf Parameter} & {\bf Mean Value} & {\bf Standard Deviation}
& {\bf Minimum} & {\bf Maximum} \\
\hline
${[Fe/H]}_{JK96}$ (dex)& -1.58(.02)$^a$ & 0.41 & -2.69$^b$ & -0.22 \\
${[Fe/H]}_{C09}$ (dex) & -1.69(.02)     & 0.41 & -2.81$^b$ & -0.33$^b$ \\
$M_V$             & 0.49(.004)     & 0.09 & 0.26      & 0.78 \\
$\mu$             & 19.13(.01)     & 0.19 & 18.37     & 19.70 \\
$d (kpc)$         & 67.31(.27)     & 5.82 & 47.21     & 87.12 \\
\hline
\end{tabular}
\end{center}
$^a$ The numbers in parentheses denote the corresponding standard errors.\\
$^b$ These very low and high metallicity values lie beyond the range of the applicability of the calibrating equations and therefore are less reliable. The lowest reliable values on the JK96 and C09 scales are ${[Fe/H]}_{JK96}=-2.25$ dex and ${[Fe/H]}_{C09}=-2.36$ dex, respectively, while the highest reliable value on the C09 scale is ${[Fe/H]}_{C09}=-0.69$ dex.\\
\end{minipage}
\end{table*}

Following Paper I, the metal abundances [Fe/H] of RR Lyrae variables pulsating in the fundamental-mode (RRab) were derived from empirical relations involving the Fourier decomposition parameters calculated in the previous section. We used the linear relation of JK96 for the RRab stars (containing the period and the Fourier phase $\varphi_{31}$). Application of the JK96 calibration requires that the RRab light curves satisfy a certain completeness and regularity criterion referred to by the authors as a compatibility test, quantified via the deviation parameter $D_m$ (details are given in JK96 and also in Paper I). Thus, we have derived $D_m$ and its standard deviation, $\sigma_{D_m}$, for all 1887 RRab stars in our sample. These values are presented in the last Column of Table 1. Following the selection process described in Paper I, we ended up with a total of 454 RRab stars that satisfy the compatibility criterion with $\langle D_m \rangle=3.45\pm1.01$ (std) and $\langle \sigma_{D_m} \rangle=1.88\pm0.54$ (std). The importance of applying the $D_m$ criterion in order to derive reliable metal abundances is thoroughly discussed in Paper I for interested readers.

\par{Table 3 summarizes the average values and associated standard deviations of the Fourier decomposition parameters for the final sample of 454 RRab stars. These values and the corresponding standard deviations are very similar to those derived for the much smaller sample of Paper I. The average value of the most important parameter, $\varphi_{31}$, for these stars was found to be $4.92\pm0.30$ (std). The corresponding value in Paper I was $4.97\pm0.30$ (std), while JK96 derived a value of 5.1 for fundamental-mode RR Lyrae stars, both results being in agreement with the present one.}

\par{The metal abundances, [Fe/H]$_{JK96}$ and [Fe/H]$_{C09}$, for the 454 RRab variables of the final sample were derived on the JK96 and Carretta et al. (2009, hereinafter C09) metallicity scales, using the appropriate calibration equations (2) and (3) of Paper I (i.e. A2 and A3, respectively). The values of $[Fe/H]$ on both scales are listed in Table 4 (Columns 2 and 3), along with their standard deviations (given in parentheses) derived from equations (4) and (5) of Paper I (i.e. A4 and A5, respectively), while statistics of the derived metallicities are presented in the first two lines of Table 5.}

\par{Equation (2) of Paper I (i.e. A2, JK96 calibration) is derived from RRab stars with metal abundances ([Fe/H]) from $-2.1$ to $+0.27$ dex, while 37 stars in our sample have metallicities below this range (-2.69 dex being the metallicity of the most metal poor star on the JK96 scale). Of these, 19 are still outliers even taking into account their $1\sigma$ error. Thus, the lowest reliable value in Table 4 on the JK96 scale is ${[Fe/H]}_{JK96}=-2.25$ dex (see Table 5). As already mentioned in Paper I, linear extrapolation of the empirical calibrating equations may lead to erroneous results, particularly since the metal poor tail of the JK96 calibrating stars is only populated by 3 objects. Our averages are bound to be affected by this problem but not significantly due to the relatively small number of outliers and their spatial distribution in a large region which doesn't affect the statistics in smaller ones. The resulting metal abundances were transformed to $[Fe/H]_{C09}$ in the C09 scale, using equation 3 of Paper I (i.e. A3), which is valid for $[Fe/H]_{JK96}$ between -2.31 and -0.68 dex. Of 435 RRab stars that lie within the validity range of equation (2) of Paper I (A2), 10 are out of the validity range of equation (3) of Paper I (A3), among them 5 being still outliers after taking into account their $1\sigma$ error. Thus, of our sample of 454 RRab stars, 430 have reliable metallicities on both JK96 and C09 scales. The lowest and highest reliable values on the C09 scale are ${[Fe/H]}_{C09}=-2.36$ dex and ${[Fe/H]}_{C09}=-0.69$ dex, respectively (see Table 5). Despite the uncertainties of the metallicities of the remaining 24 objects, we decided to keep them in our final sample, since they are not expected to affect our results due to their small number ($\simeq 5\%$ of the total).}

\subsection{Distribution of metal abundances}

The distributions of the metal abundances of the RRab stars on the C09 and JK96 scales are shown in Fig. 1 (with light-grey and dashed bars, respectively), The corresponding averages (listed in Table 5) are $\langle [Fe/H]_{C09} \rangle =-1.69\pm0.41$ dex (std, while the standard error is 0.02 dex) and $\langle [Fe/H]_{JK96} \rangle=-1.58\pm0.41$ dex (std, the standard error being 0.02 dex). The distributions appear to be different from those derived in Paper I (dark grey bars in Fig. 1 for the C09 scale), where only the RRab stars from the central region of the SMC were included (fig. 4 of Paper I). The extended region of the SMC seems to include more metal poor objects. This can be seen in Table 6, where a comparison between RRab populations of different metal abundances in the central bar region (Paper I) and the extended area (Paper II) of the SMC is shown to evaluate the difference between the two distributions. Two extreme subsamples are used, i.e. stars with $[Fe/H]_{C09} \le -2.0$ (metal poor stars) and those with $[Fe/H]_{C09} \ge -1.4$ (metal rich stars). The percentage of metal poor stars in the samples of Paper I and the present work is $19\%$ and $24\%$, respectively, while $0.62\pm0.20$ and $1.08\pm0.15$ are the corresponding ratios of the number of these stars to the number of the metal rich ones. However, this issue will be further discussed in Section 4. It should be noted that a Kolmogorov-Smirnov (K-S) test showed that the distributions of $[Fe/H]_{C09}$ in Fig. 1 (light and dark grey bars) are identical while a $\chi^2$ test indicates that this hypothesis is true at a low level of significance (i.e. $0.1$).

\begin{figure}
\begin{center}
\leavevmode \epsfxsize=95mm \epsffile{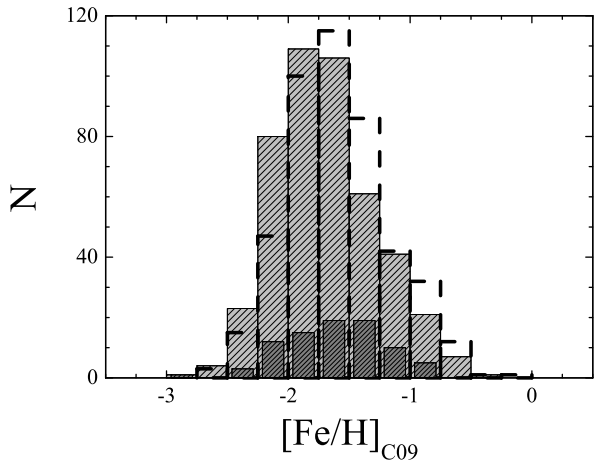}
\end{center}
\caption{Distribution of the metal abundances of 454 RRab stars (light grey bars). [Fe/H] is calculated on the new scale of C09. The dash line corresponds to the distribution of the metallicities of these RRab stars on the JK96 scale, while the dark grey bars correspond to the distribution of the metal abundances of 84 RRab stars in the central bar region of the SMC (Paper I) on the C09 scale.}
\end{figure}

\subsection{Comparison with other studies}

Very recently, HGDJ have reported an average metallicity of $-1.42$ dex from RRab stars that were detected from OGLE-III in the SMC, which should be compared with our average value of $-1.58\pm0.02$ dex (standard error) on the JK96 scale. Inspection of the corresponding distributions (fig. 6 of HGDJ and the dashed bars in our Fig. 1) indicates that the high-metallicity bins in the HGDJ distribution are more populated than the low-metallicity bins compared to ours. A comparison between the metal abundances (on the JK96 scale) of 453 individual RRab stars which are common in HGDJ's and our final sample is shown in the upper panel of Fig. 2, where a 1:1 line is also plotted. Systematic offsets are evident since HGDJ's metallicities are increasingly higher than ours for metal poorer stars while the opposite trend (although weakened) appears for the metal rich ones, the average absolute difference ($<|\Delta[Fe/H]_{JK96}|>$) being $0.27\pm0.21$ dex (std). Furthermore, the average $< \frac{|\Delta[Fe/H]_{JK96}|}{\sigma_{[Fe/H]_{JK96}}} > = 2.27\pm1.70$ (std), where $\sigma_{[Fe/H]_{JK96}}$ is the error (std) of our estimate, is a characteristic of large deviations ($|\Delta[Fe/H]|<1\sigma$ for only $27\%$ of the 453 RRab stars). In the lowel panel of Fig. 2 the difference between the present and the HGDJ's metallicity estimates are plotted as a function of our results for each of the 453 stars. The equation describing a linear fitting is $\Delta [Fe/H]_{JK96}^{present-HGDJ} = (0.70\pm0.04) + (0.55\pm0.02)[Fe/H]_{JK96}^{present}$. It should be noted that HGDJ used I-band light curves of 1831 variables with presumably reliable metallicities and the calibration equation of Smolec (2005), which is similar to the JK96 one but is applied on I-band Fourier decomposition parameters. However, this calibration equation is based on a very small sample of 28 field RRab stars and it is not accompanied by any compatibility condition, while the JK96 calibration is based on a larger sample of 84 carefully selected RRab stars along with a strict criterion on the quality of the light curves. HGDJ repeated their analysis following the method of DS10 and found similar results to their first method, although the corresponding average metallicity was slightly lower, i.e. $-1.53$ dex. As expected, the systematic trends shown in the upper panels of Fig. 2 (comparison between HGDJ's and the present work) and fig. 6 of Paper I (comparison between DS10's and our method) are identical. As thoroughly discussed in Paper I, a comparison between metallicities derived from V- and I-band light curves reveals systematic differences, mainly in the individual values and their distributions, with at least part of the discrepancies stemming from the lack of a compatibility test for the light curves used.

\begin{figure}
\begin{center}
\leavevmode \epsfxsize=90mm \epsffile{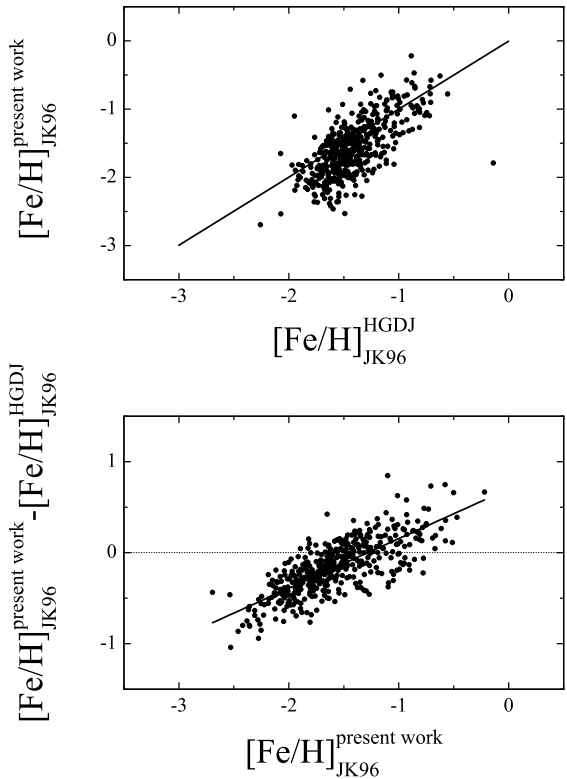}
\end{center}
\caption{Upper panel: Comparison between the metal abundances (on the JK96 scale) of 453 RRab stars in the SMC which were derived using the HGDJ's approach and the present one, i.e. based on Fourier decomposition of I- and V-band light curves, respectively (a 1:1 line is also plotted). Lower panel: Difference between the two estimates of the metal abundance for each RRab star as a function of the metallicities derived in the present work (a linear fitting is also plotted).}
\end{figure}

\begin{figure}
\begin{center}
\leavevmode \epsfxsize=95mm \epsffile{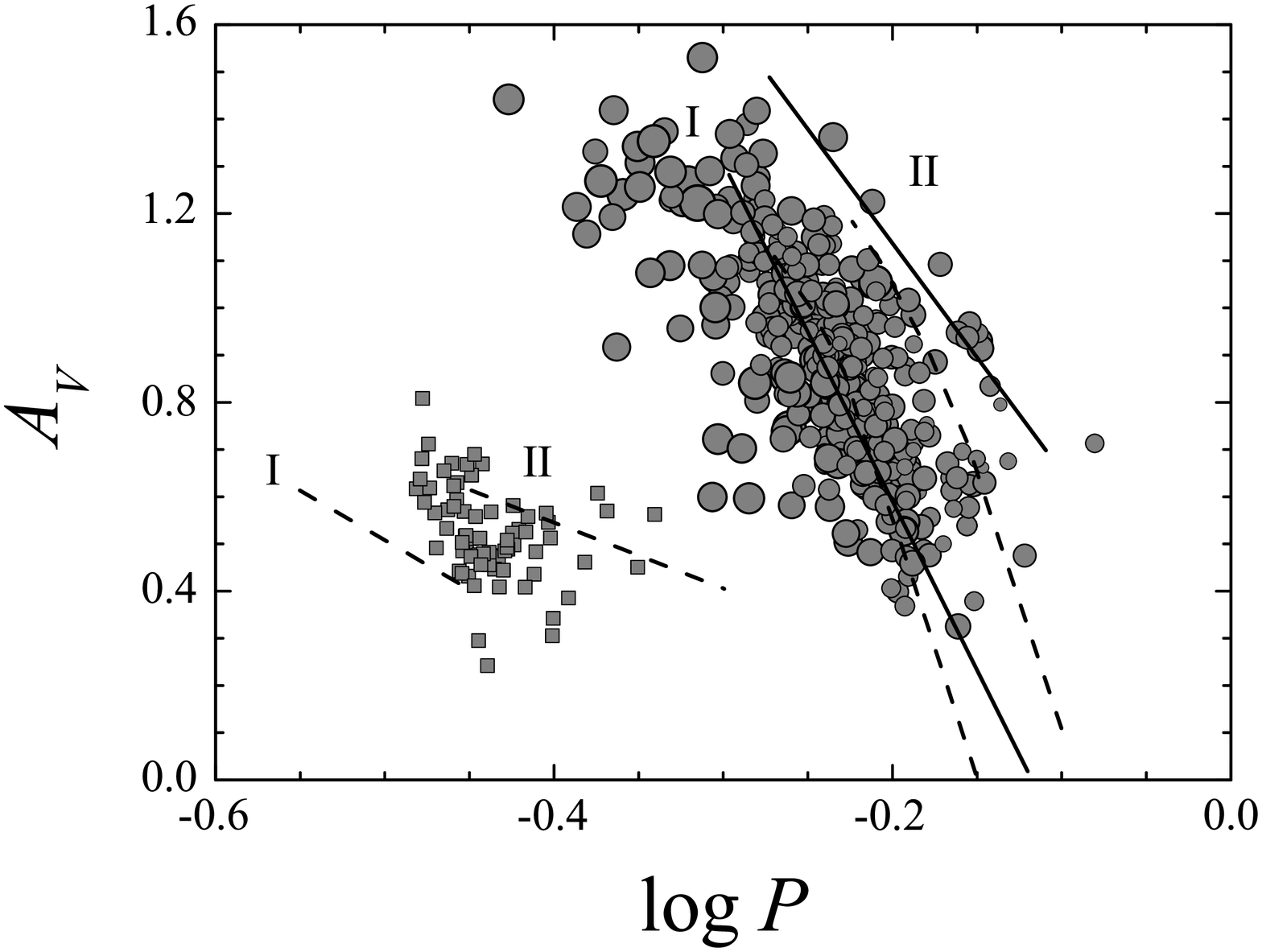}
\\
\leavevmode \epsfxsize=95mm \epsffile{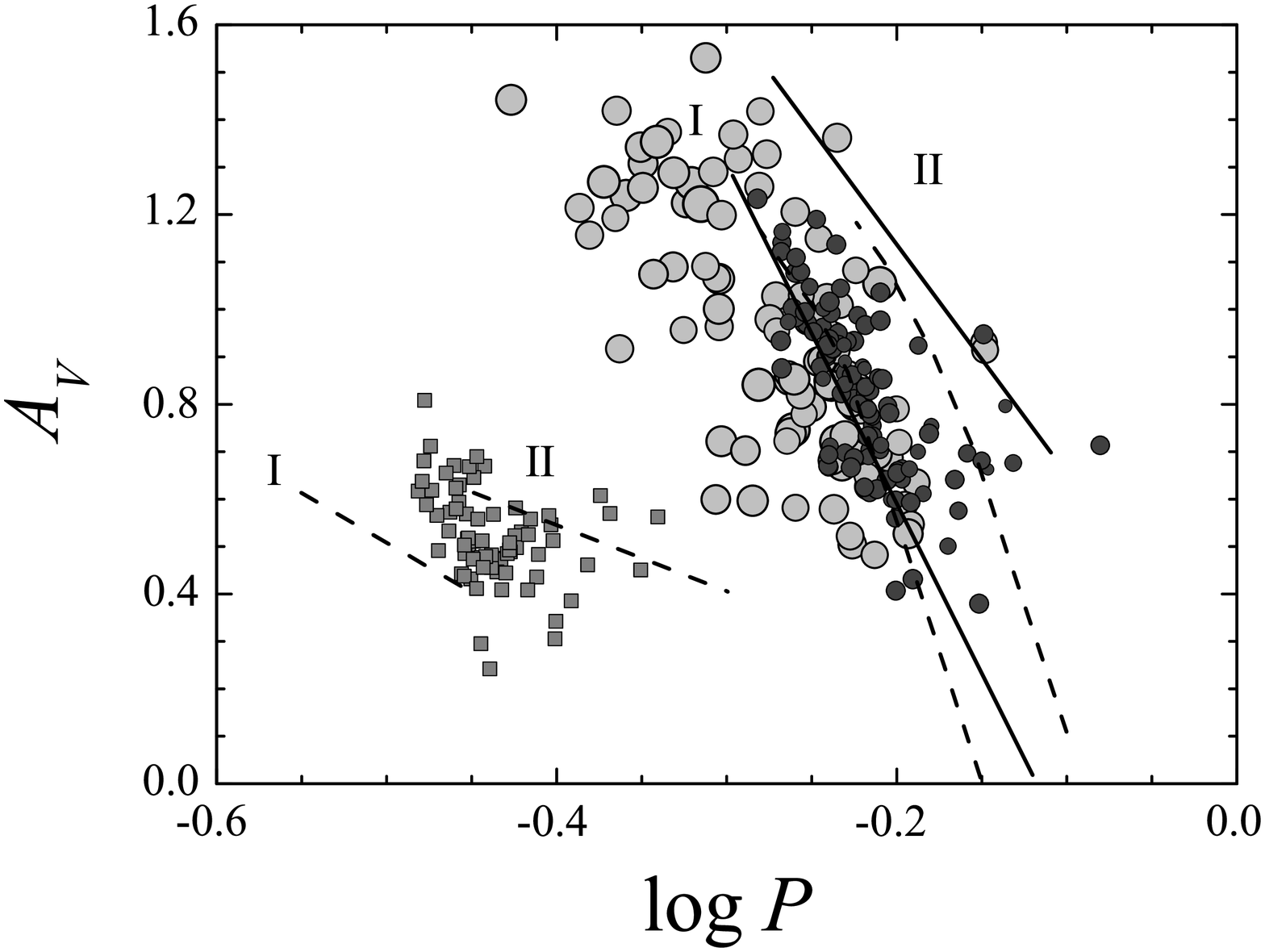}
\end{center}
\caption{Upper panel: Bailey diagram (amplitude as a function of $\log P$). The grey circles correspond to RRab stars, while the increasing size denotes increasing metal abundance. The positions of the RRc stars on the Bailey diagram are also shown with grey squares for completeness reasons. The solid curves represent the ridge lines for RRab stars on OoI and OoII clusters, according to Clement \& Rowe. The dashed lines correspond to the loci of RRab and RRc stars of the M3 cluster having OoI properties and a number of evolved stars off the ZAHB that mimic the OoII behavior, according to Cacciari, Corwin \& Carney.
Lower panel: Bailey diagram for two extreme subsets. The circles correspond to RRab stars while the squares correspond to RRc stars. The increasing size of the circles denotes increasing metal abundance. The light grey symbols represent RRab stars with metal abundances greater than -1.40 dex on the C09 scale, while the dark grey symbols represent those RRab stars with metal abundances less than -2.00 dex on the C09 scale.}
\end{figure}

\begin{table}
\begin{minipage}{85mm}
\begin{center}\scriptsize
\caption{Comparison between RRab populations of different metal abundances in the central bar region (Paper I) and the extended area (present work: Paper II) of the SMC.}
\begin{tabular}{@{}ccc@{}}
\hline
Statistics          & Paper I & Paper II \\
\hline
Sample size                              \\
$N$                 & 84      & 454      \\
\\
RRab stars with $[Fe/H]_{C09} \le -2.0$ dex \\
$N_{poor}$          & 16      & 108      \\
$N_{poor}\%$        & 19$\%$  & 24$\%$   \\
\\
RRab stars with $[Fe/H]_{C09} \ge -1.4$ dex \\
$N_{rich}$          & 26      & 100      \\
$N_{rich}\%$        & 31$\%$  & 22$\%$   \\
\\
$N_{poor}/N_{rich}$ & $0.62\pm0.20$    & $1.08\pm0.15$     \\
\hline
\end{tabular}
\end{center}
\end{minipage}
\end{table}

\subsection{The Bailey diagram}

The theoretical models and the expected behavior of the RR Lyrae stars on the Bailey diagram (as well as its usage, e.g. Soszy\'{n}ski et al. 2003) are thoroughly discussed in Paper I.

\par{The upper panel of Fig. 3 shows the loci of the fundamental-mode RR Lyrae variables (grey circles) in our sample on the Bailey diagram, while in the lower panel we have overplotted the distribution on the Bailey diagram of two extreme subsets, i.e. RRab stars with metal abundances greater than -1.40 dex (light grey circles) and those with metal abundances less than -2.00 dex on the C09 scale (dark grey circles), in order to facilitate the comparison. In both diagrams the positions of the first overtone RR Lyrae variables (grey squares) are shown for completeness reasons. Furthermore, we have overlayed the standard ridge lines for RRab stars in the Galactic globular cluster M3 (prototype OoI) and $\omega$ Centauri (OoII) from Clement \& Rowe (2000) with solid lines, while the dashed lines denote the loci of of the bona fide regular (OoI) and evolved (falling closer to OoII line) RRab stars in M3 from Cacciari, Corwin \& Carney (2005). In both cases, OoI and OoII (Oosterhoff 1939) are denoted by the left and the right curves, respectively.}

\par{Our basic results from the RR Lyrae stars of the central region of the SMC in Paper I are confirmed from the present larger sample. The metal rich RRab stars of our sample seem to lie closer to the OoI curves (M3 non-evolved sequence in the interpretation of Cacciari, Corwin \& Carney), while the more metal poor objects extend slightly towards the evolved sequence (OoII curves). Yet, a large number of the latter is located close to the OoI curve. The bulk of the RR Lyrae variables with intermediate metallicities are located in the region between the two Oosterhoff curves, possibly constituting an Oo-intermediate population. Thus, the location of a particular RRab variable on the Bailey diagram seems to be affected both by metal abundance and evolution off the ZAHB.}

\section{Metallicity gradient}

\begin{figure*}
\begin{center}
\leavevmode \epsfxsize=18cm \epsffile{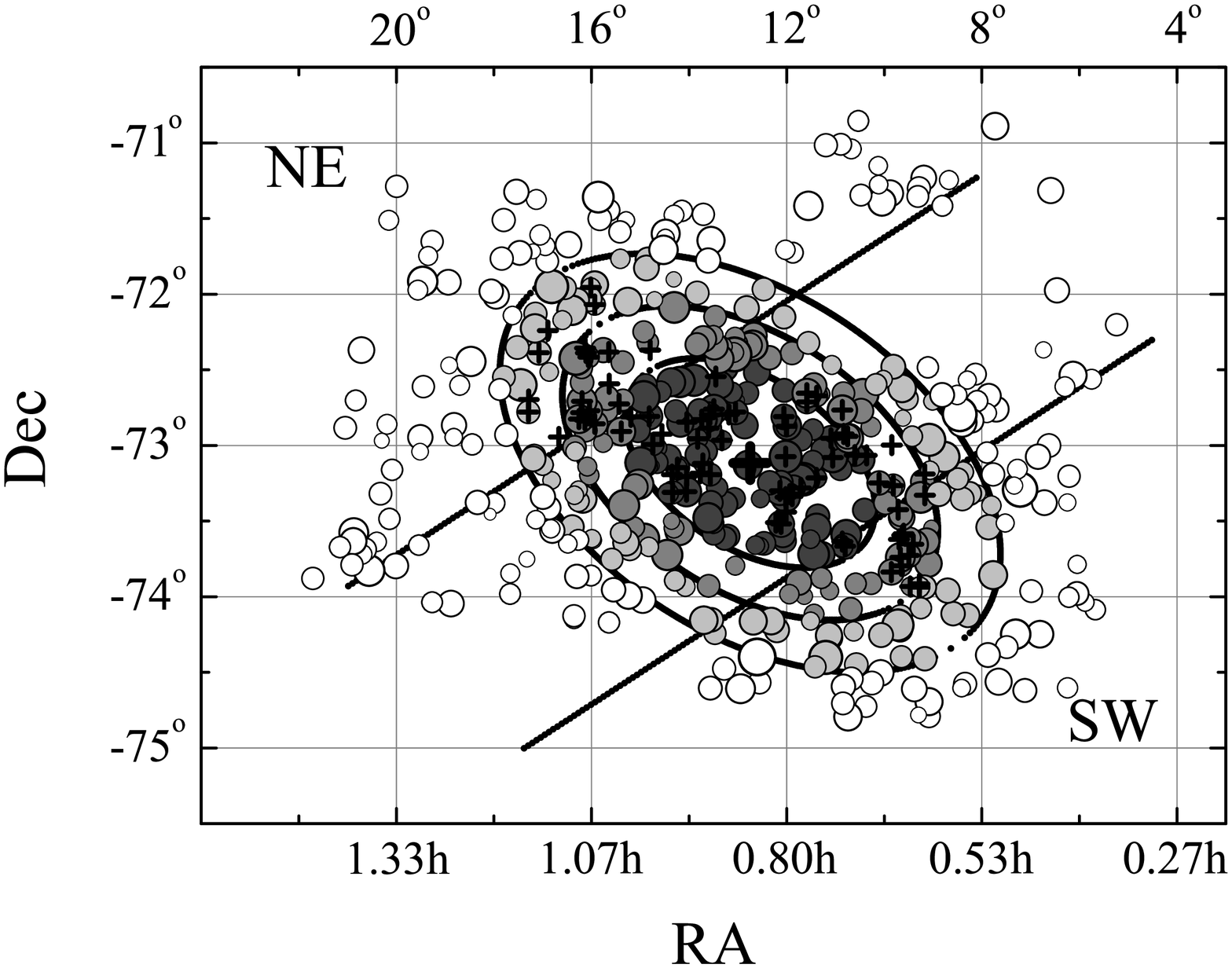}
\\
\leavevmode \epsfxsize=15cm \epsffile{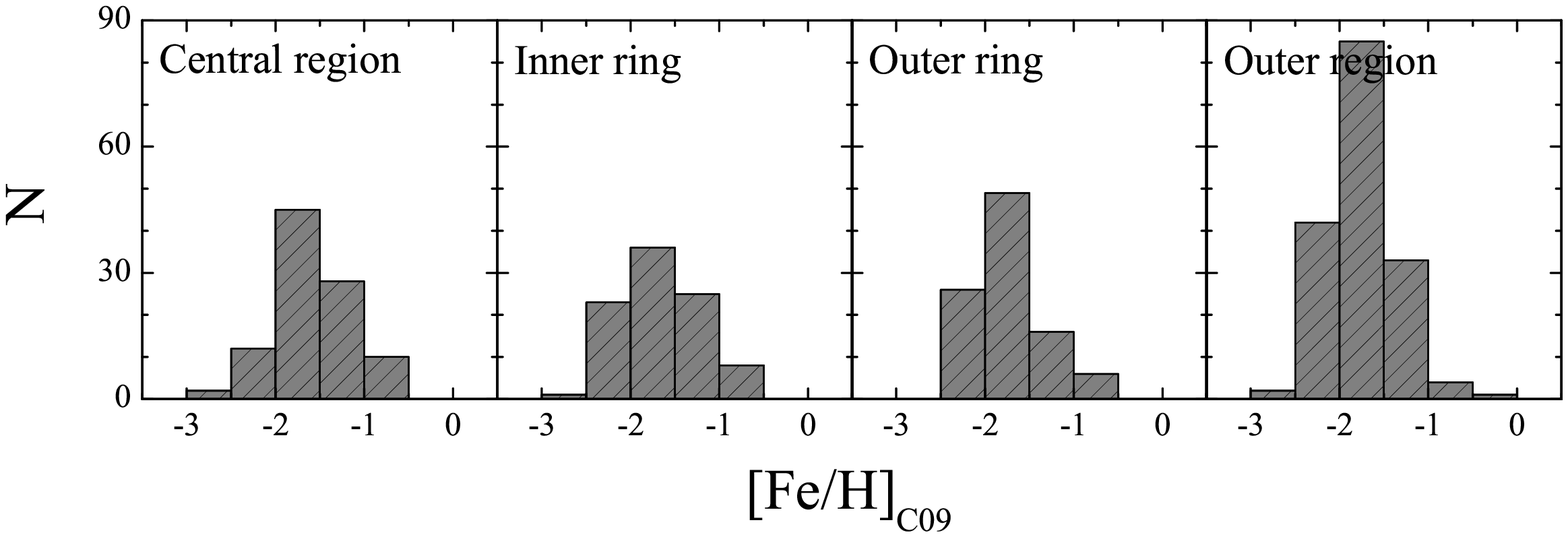}
\end{center}
\caption{Distribution of the 454 fundamental-mode RR Lyrae stars in four regions separated with ellipses, centered on the SMC dynamical center and based on the isopleths derived by Gonidakis et al. Dark-grey, grey, light-grey and open circles correspond to stars lying in the central region, an inner ring, an outer ring and the outer region, respectively. The two parallel black lines (which are perpendicular to the ellipses major axes) are used to separate the RR Lyrae variables of the north-eastern (NE) and south-western (SW) regions of the SMC from the bulk of our sample. The increasing symbol size denotes increasing metallicity. The thick big cross indicates the position of the SMC dynamical center, while the smaller crosses correspond to the 84 RRab stars from the sample of the central bar region in Paper I. The histograms correspond to the distributions of the metal abundances ($[Fe/H]_{C09}$) in the four regions mentioned above.}
\end{figure*}

\begin{table*}
\begin{minipage}{180mm}
\begin{center}\scriptsize
\caption{Comparison between RRab populations in four areas of the SMC, separated by ellipses.}
\begin{tabular}{@{}ccccc@{}}
\hline
Area of                        & central        & inner          & outer          & outer          \\
the SMC                        & region         & ring           & ring           & region         \\
\hline
Sample size                                                                                        \\
$N$                            & 97             & 93             & 97             & 167            \\
$\langle [Fe/H]_{C09} \rangle (dex) \pm std (se)$ & $-1.60\pm0.42(0.04)$ & $-1.65\pm0.44(0.05)$ & $-1.71\pm0.40(0.04)$ & $-1.76\pm0.39(0.03)$ \\
$1\sigma$ line-of-sight depth (kpc)  & $6.38\pm0.60$  & $5.55\pm0.12$  & $5.23\pm0.14$  & $4.46\pm0.24$  \\
\\
Stars with $[Fe/H]_{C09} \le -2.0$ dex                                                             \\
$N_{poor}$                     & 14             & 24             & 26             & 44             \\
$N_{poor}\%$                   & 14$\%$         & 26$\%$         & 27$\%$         & 26$\%$         \\
$1\sigma$ line-of-sight depth (kpc)  & $6.96\pm0.22$  & $7.59\pm0.13$  & $5.12\pm0.14$  & $4.97\pm0.25$  \\
\\
Stars with $[Fe/H]_{C09} \ge -1.4$ dex                                                             \\
$N_{rich}$                     & 30             & 24             & 17             & 29             \\
$N_{rich}\%$                   & 31$\%$         & 26$\%$         & 18$\%$         & 17$\%$         \\
$1\sigma$ line-of-sight depth (kpc)  & $5.93\pm1.34$  & $4.26\pm0.13$  & $5.22\pm0.17$  & $4.62\pm0.13$  \\
\\
$N_{poor}/N_{rich}$            & $0.47\pm0.15$  & $1.00\pm0.29$  & $1.53\pm0.48$  & $1.52\pm0.36$  \\
\hline
\end{tabular}
\end{center}
\end{minipage}
\end{table*}

In the central bar region of the SMC a metallicity gradient was neither expected nor observable (see section 4.5 of Paper I). In the present study, the area covered by OGLE-III is extended to a radius of $\simeq4^{\circ}$ on average (approximately 14 square degrees). The distributions of metallicities of the RRab stars in the central region of the SMC (fig. 4 of Paper I and dark grey bars of Fig. 1 in the present work) and the extended area (light grey bars of Fig. 1), imply the possible existence of a metallicity gradient, since there is a higher percentage of metal poor objects in the extended sample. In order to examine the genuineness of such an effect, we derived the average metal abundances in four regions, defined by ellipses based on the isopleths of Gonidakis et al. (2009, hereinafter G09) and centered on the SMC dynamical center (DC). A central region, an inner ring, an outer ring and the outer region are shown in the upper panel of Fig. 4, overlayed on the map of the 454 RRab stars. The distributions of the metal abundances in the four regions are shown in the lower panels of Fig. 4. The corresponding average metallicities on the C09 scale ($\langle [Fe/H]_{C09} \rangle$) with increasing angular distance from the SMC DC are $-1.60\pm0.42$ dex, $-1.65\pm0.44$ dex, $-1.71\pm0.40$ dex and $-1.76\pm0.39$ dex, the errors being standard deviations (the corresponding standard errors being $\pm0.04$, $\pm0.05$, $\pm0.04$ and $\pm0.03$). Although these average values are identical within their standard deviations (which is not the case for the standard errors), a small but systematic decrease of the average metal abundance is detected with increasing distance from the SMC DC. Furthermore, inspection of the distributions of the metallicities in the four regions (lower panels of Fig. 4) reveals an increasing relative surplus of high metallicity objects towards the inner regions. This can be also seen in Table 7, where a comparison between two extreme subsets of RRab stars in each area is presented. We have examined two extreme populations of stars with $[Fe/H]_{C09} \le -2.0$ dex (metal poor) and $[Fe/H]_{C09} \ge -1.4$ dex (metal rich) as well as their number ratio. The percentage of the metal poor stars increases with angular distance from the SMC dynamical center while the opposite occurs for the metal rich stars. Their ratio $N_{poor}/N_{rich}$ also shows a similar increase. Furthermore, a $\chi^2$ test indicates that the hypothesis of all samples of non-adjacent regions sharing the same origin is true only below very low levels of significance ($0.1-0.5$). A K-S test for the two extreme regions (inner and outer region) indicates that this hypothesis is true on a 0.05 level of significance.

\par{On the other hand, it should be noted that the RRab population in the inner regions contains low metallicity (as well as some high metallicity) stars whose actual (de-projected) distance from the SMC DC is greater than the corresponding projected distance which was used in the analysis above. These stars lie behind or in front of the SMC central region and their exclusion from the sample of RRab stars in the inner regions would lead to slightly higher average metallicities, leading thus to a stronger evidence of a metallicity gradient existing in the SMC. For example, the innermost region contains 97 RRab stars with an average distance of $68\pm7$ kpc (std) from us, the larger and shorter distance being $87$ kpc and $52$ kpc, respectively. Applying a cutoff and keeping stars with a distance diverging less than $\pm1.5\sigma$ from the average one would lead to a sample of 83 stars with an average metallicity of $-1.57\pm0.40$ dex (std) on the C09 scale. A $\pm1\sigma$ cutoff would lead to a similar metallicity of $-1.57\pm0.41$ dex (std) on the C09 scale for 72 stars. Both results are slightly higher than the value of $-1.60\pm0.42$ dex that was found earlier.}

\par{Thus, another approach was followed using the real (de-projected) distances ($d_{DC}$) of the RRab stars from the SMC DC. For this purpose, we used their distances (d) from us (as they were derived with the method described in Section 5), their angular distances ($\omega$) from the SMC DC (using the coordinates RA and Dec of each star and those of the DC, i.e. ($\alpha$, $\delta$) and ($\alpha_o = 0^h\;51^{min}$, $\delta_o = -73^{\circ}\;7^{\prime}$), respectively) and adopting the average distance of the SMC ($D=67.31\pm5.82$ kpc, being the average of the individual distances of the RRab stars from us, see Section 5.3) as the distance of the DC. The corresponding equations are described below.
\begin{equation}
{d_{DC} = \sqrt{ d^2 + D^2 - 2 d D \cos{\omega}}}
\end{equation}
where
\begin{equation}
{\cos{\omega} = \sin{\delta}\sin{\delta_o} + \cos{\delta}\cos{\delta_o}\cos{(\alpha - \alpha_o)}}
\end{equation}
The average metallicities ($\langle [Fe/H]_{C09} \rangle$ on the C09 scale) were derived for 10 groups of stars with increasing average distance from the SMC DC ($\langle d_{DC} \rangle$), from 0 to 20 kpc, the step (binning) being 2 kpc. This is shown in the upper panel of Fig. 5, where a small but clear decrease is found and can be well fitted by a linear relation between $\langle [Fe/H]_{C09} \rangle$ and $\langle d_{DC} \rangle$, as described by the following equation (the last two poorly populated groups, corresponding to the open circles in Fig. 5, were excluded from the fit):
\begin{equation}
{\langle [Fe/H]_{C09} \rangle = -(0.013\pm0.007)\langle d_{DC} \rangle - (1.624\pm0.062)}
\end{equation}
where $r=0.623$ is the correlation coefficient and $\sigma=0.210$ the corresponding standard deviation.}

\begin{figure}
\begin{center}
\leavevmode
\epsfxsize=80mm \epsffile{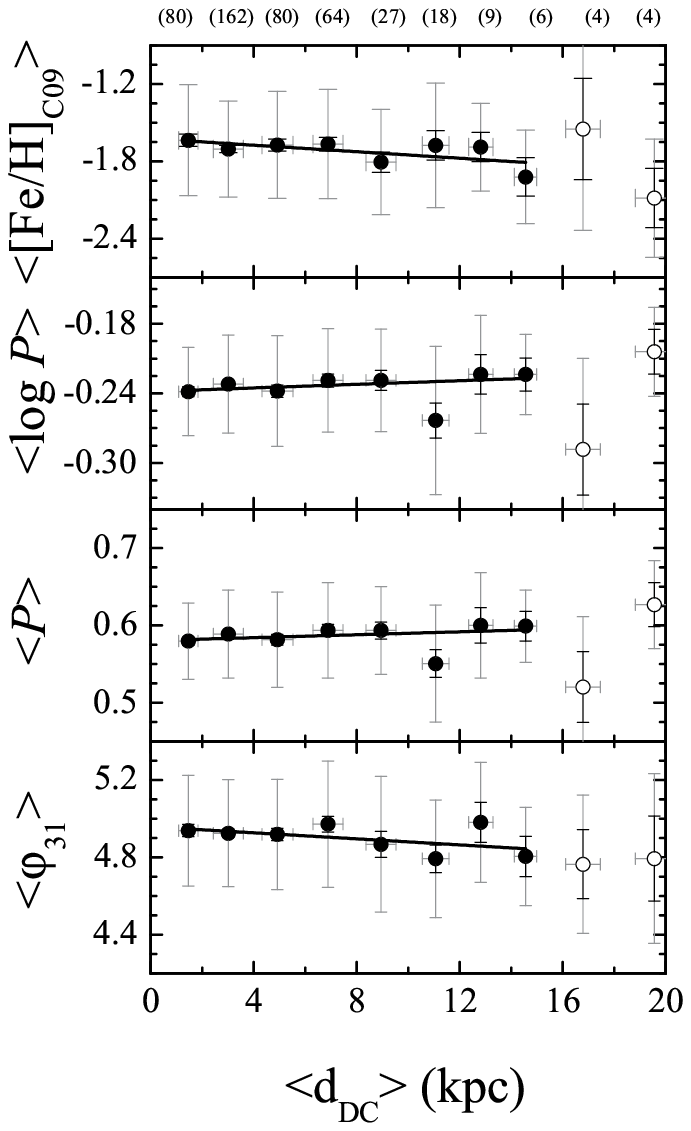}
\end{center}
\caption{Average metallicities, $\log P$, $P$ and $\varphi_{31}$ of 10 groups of RRab stars as a function of their average distance (in kpc) from the SMC dynamical center. The grey and black error bars are standard deviations and standard errors, respectively, while the solid lines represent linear fittings. The open circles denote poorly populated groups that were excluded from the fittings. The numbers in parentheses denote the number of stars in each group.}
\end{figure}

\begin{figure}
\begin{center}
\leavevmode \epsfxsize=70mm \epsffile{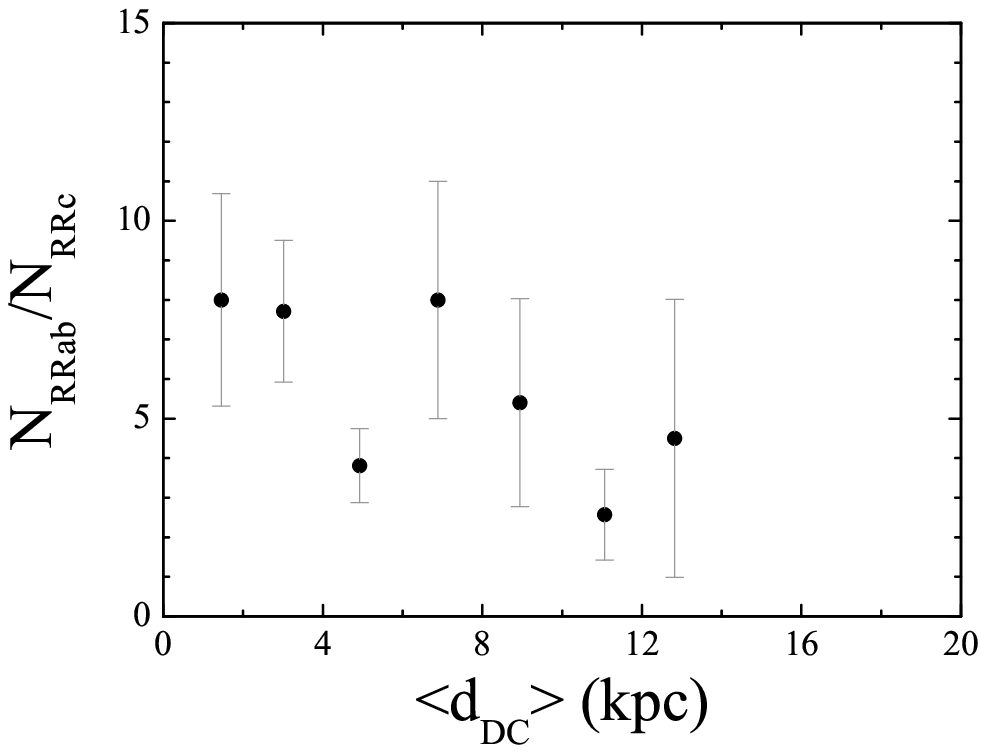}
\\
\leavevmode \epsfxsize=67mm \epsffile{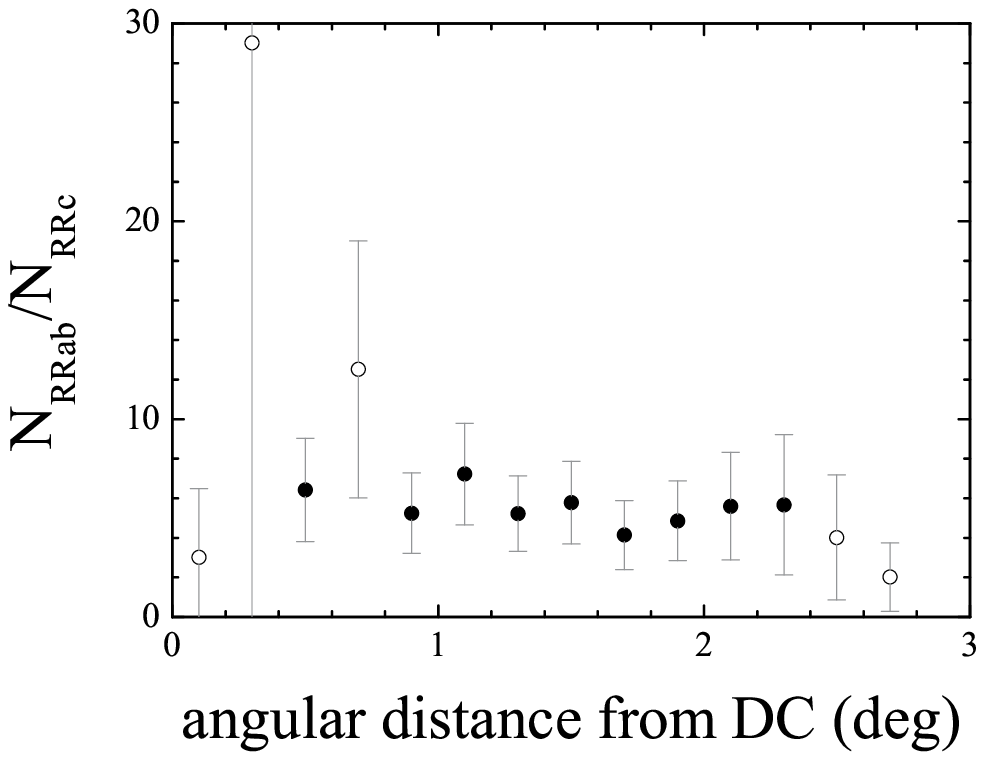}
\end{center}
\caption{Upper panel: Ratio of the number of RRab ($N_{RRab}$) and the corresponding RRc ($N_{RRc}$) stars for groups of RRab stars (see Fig. 5) as a function of their average distance from the SMC dynamical center (applied for groups were both populations were present), based on the sample of the 454 RRab and 74 RRc stars which are compatible with the criteria based on $D_m$ and $\sigma_{\varphi_{31}}$ parameters, respectively. Lower panel: Same ratio ($N_{RRab}/N_{RRc}$) for groups of stars with increasing angular distance from the SMC dynamical center, based on all 1887 RRab and 169 RRc stars which were detected from OGLE-III.}
\end{figure}

\subsection{Discussion of errors and biases}

As already discussed in Paper I, selection effects may be at play; photometric crowding would make larger amplitude (hence more metal rich) RR Lyrae stars (see Bailey diagram in Fig. 3) easier to detect and with higher quality light curves (hence smaller $\sigma_{D_m}$ and $D_m$) in the inner region than lower metallicity RR Lyrae stars in the same area. A detailed study of the line-of-sight depth variations in Section 5.5 reveals such a possible lack of metal poor RRab stars in the innermost regions of the SMC (below 1 deg), although their actual distances from the DC may be just as large. Since RRc stars are on average of lower amplitudes than the RRab stars, such a selection effect would also imply an increase of the ratio $N_{RRab}/N_{RRc}$ close to the SMC center, where $N_{RRab}$ and $N_{RRc}$ are the populations of RRab and RRc stars, respectively, in the same area (Feast, Abedigamba \& Whitelock 2010, hereinafter FAW10). In order to examine this possibility we derived this ratio for 7 (most populated) of the 10 regions mentioned above (provided that both populations were present in each region). For this purpose, we used the sample of 454 RRab stars whose metal abundances and thus their distances were derived with accuracy (see Sections 3 and 5). Concerning the RRc stars, although they were excluded from further analysis in this project (see discussion in Section 1 and in Paper I), we derived their distances with the same method that we adopted for the RRab stars (see Section 5 and equations 1 and 2 above) and their metallicities following the technique described in Paper I (equations 6-10 and 12 there or A6-10 and A12 in the Appendix). These values are not suitable for accurate chemical and structural analysis of the SMC, however they could be used for the purpose of the statistical test that is described here. Of the 169 RRc stars, 74 objects were compatible with the selection criterion based on their $\sigma_{\varphi_{31}}$ (see Paper I for details) and were used for this analysis. The results are given in the upper panel of Fig. 6, where a decrease in the ratio $N_{RRab}/N_{RRc}$ with increasing distance from the SMC DC is shown. Thus, the metallicity gradient found above is possibly affected by selection effects related to the OGLE observations (undetected metal poor RRab stars of low amplitude in the central region) and the selection criteria which led to our final sample (metal poor RRab stars of higher $\sigma_{D_m}$ and $D_m$ being excluded) as well. It should be noted that the excluded metal poor RRab stars with small angular distances from the DC could be either of small radial distances from the DC, leading thus to an overestimated metallicity gradient, or of larger distances, therefore underestimating the gradient. The ratio $N_{RRab}/N_{RRc}$ is also affected by these effects; undetected and excluded (high $\sigma_{\varphi_{31}}$) RRc stars would imply its increase in the inner regions whereas undetected and excluded metal poor RRab stars close to the DC would lead to its decrease. We derived a similar ratio from the total sample of 1887 RRab and 169 RRc stars that were detected from OGLE-III. However, since the actual distances from the SMC DC could not be derived for all these stars with accuracy, we used their angular distances, although the groups with small angular distances also include stars with high actual distances. The corresponding diagram is shown in the lower panel of Fig. 6. The slope of a linear fitting is $-0.81\pm0.60$ (where the open circles corresponding to poorly populated groups were excluded), i.e. zero within its $1.3 \sigma$ error.

\par{For a further examination of possible biases originating from our sample, we investigated whether the metallicity gradient would be detectable either by adopting a different binning or by excluding certain points from the fitting procedure, as it shown in Table 8. Thus, we changed the binning from 2.0 kpc (10 groups of stars) to 1.5 kpc and 2.5 kpc, leading to 14 and 8 groups of stars, respectively. In all cases, the number of stars (N) of each group is listed in Table 8, while the last groups in each case (marked with an asterisk in Table 8) were excluded from the analysis due to very low statistics. We performed linear fittings using all the remaining statistically reliable groups or by excluding the first or/and the last of them in each case. Those included in each fitting and the corresponding slope, as well, are listed in the last two columns of Table 8. Several conclusions may be derived from these tests. To start with, the metallicity gradient is independent of the adopted binning. Furthermore, it seems to be partially based on the last group in each case, i.e. the $10^{th}$, $8^{th}$ and $6^{th}$ for binning of 1.5, 2.0 and 2.5 kpc, respectively. On the contrary, the inclusion of the first group isn't crucial. However, any undetected metal poor stars in the innermost regions of the SMC could weaken the gradient. Unfortunately, the number of possible missing stars could not be securely evaluated (i.e. without the danger of including non-existing objects). At any rate, exclusion of certain groups (i.e. also the first or/and the last of the reliable ones in each case) would also suffer from biases and selection effects towards the opposite direction, i.e. the "washing out" of the gradient.}

\par{As a final test, relaxing the $D_m$ criterion, i.e. by selecting stars with $\sigma_{D_m} \leq 4$ and $D_m - \sigma_{D_m} \leq 4$, would led us to a larger sample of 714 objects, although their metal abundances and distances, as well, would be beyond the limits of reliability ($<D_m>=4.08\pm1.31$, $D_m^{max}=7.46\pm3.91$, $<\sigma_{D_m}>=2.07\pm0.73$, the errors being std). In this case (e.g. by adopting a binning of 2.0 kpc), the metallicity gradient would be hardly distinguishable, i.e. $-0.006\pm0.005$ dex/kpc, its significance though being strongly questionable.}

\begin{table*}
\begin{minipage}{180mm}
\begin{center}\scriptsize
\caption{Investigation of the existence of a metallicity gradient by linear fittings on average metallicities of groups of RRab stars versus their corresponding average de-projected distances from the SMC DC, using different groupings (binning).}
\begin{tabular}{@{}cccccccccccccccccc@{}}
\hline
binning&group  &  1&  2&  3&  4&  5&  6&   7 &   8 &   9 &  10 &  11 &  12 &  13 &  14 &fitted& slope (dex/kpc)\\
\hline
1.5kpc&N stars& 40&120&109& 53& 51& 28&  14 &  16 &   7 &   7 &3$^*$&2$^*$&1$^*$&3$^*$& 1-10 &$-0.013\pm0.006$\\
      &       &   &   &   &   &   &   &     &     &     &     &     &     &     &     & 2-10 &$-0.013\pm0.007$\\
      &       &   &   &   &   &   &   &     &     &     &     &     &     &     &     & 1- 9 &$-0.008\pm0.006$\\
      &       &   &   &   &   &   &   &     &     &     &     &     &     &     &     & 2- 9 &$-0.006\pm0.008$\\
\\
2.0kpc&N stars& 80&162& 80& 64& 27& 18&   9 &   6 &4$^*$&4$^*$&  -  &  -  &  -  &  -  & 1- 8 &$-0.013\pm0.007$\\
      &       &   &   &   &   &   &   &     &     &     &     &     &     &     &     & 2- 8 &$-0.012\pm0.008$\\
      &       &   &   &   &   &   &   &     &     &     &     &     &     &     &     & 1- 7 &$-0.004\pm0.005$\\
      &       &   &   &   &   &   &   &     &     &     &     &     &     &     &     & 2- 7 &$-0.001\pm0.006$\\
\\
2.5kpc&N stars&118&170& 85& 40& 21& 11&4$^*$&5$^*$&  -  &  -  &  -  &  -  &  -  &  -  & 1- 6 &$-0.012\pm0.007$\\
      &       &   &   &   &   &   &   &     &     &     &     &     &     &     &     & 2- 6 &$-0.012\pm0.010$\\
      &       &   &   &   &   &   &   &     &     &     &     &     &     &     &     & 1- 5 &$-0.004\pm0.009$\\
      &       &   &   &   &   &   &   &     &     &     &     &     &     &     &     & 2- 5 &$+0.001\pm0.013$\\
\hline
\end{tabular}
\end{center}
$^*$ These points were excluded from linear fitting due to very low statistics.
\end{minipage}
\end{table*}

\subsection{Comparison with other investigations in the SMC}

HGDJ investigated the existence of a metallicity gradient in the SCM using RRab stars, whose I-band light curves were decomposed using Fourier analysis independently of their quality, and they reported an estimate of $0.00\pm0.06$ dex/kpc. Their study was based on projected distances of these stars from the SMC center. In order to have comparable results, we used their metal abundances along with our de-projected distances from the SMC DC for 453 common RRab stars and performed a linear fitting on the averages of the 8 groups of stars which were described earlier and were shown in Fig. 5 (adopted binning: 2.0 kpc). The resulting slope was $-0.006\pm0.005$ dex/kpc. For consistency reasons, we redefined the de-projected distances of these stars using HGDJ's metallicities and our method described in the present work. New groups of stars were selected with the same binning of 2.0 kpc. The slope of the corresponding linear fitting was reversed but remained almost zero within its error, i.e. $+0.006\pm0.005$ dex/kpc.

\par{As already discussed in Section 3.2, HGDJ's metal abundances show systematic discrepancies, when compared to ours, i.e. they are increasingly higher than ours for metal poorer stars and slightly lower for metal richer ones. Such offsets would tend to eliminate any metallicity gradient. Thus, the estimates mentioned earlier from HGDJ's metallicities are expected to imply a hardly detectable or undetectable gradient.}

\subsection{Comparison with the LMC}

\par{Interestingly, FAW10 have found similar results to ours for the LMC RR Lyrae variables. These authors used spectroscopic values for metallicities along with period - [Fe/H] relations and suggest the existence of a radial metallicity gradient in this galaxy, detected from the RRab population, for distances up to 6 kpc from the center. The metallicity gradients in both Magellanic Clouds would be consistent with the theory of galactic evolution by the gradual collapse of a gas cloud. It should be noted though that FAW10 interpret their metallicity gradient as a result of a $\log P$ gradient, noting that the latter could also be explained in terms other than metallicity, together with an age gradient and a nearly constant mean metallicity. Since our metallicites are based on a $[Fe/H] = f(P , \varphi_{31} )$ relation, we derived the average $\log P$, $P$ and $\varphi_{31}$ for our ten groups of stars (described above) with increasing average distance from the SMC DC, as it is shown in the lower panels of Fig. 5. The corresponding linear fittings are expressed through the equations below:
\begin{equation}
{\langle \log P \rangle = (0.001\pm0.001)\langle d_{DC} \rangle - (0.238\pm0.007)}
\end{equation}
\begin{equation}
{\langle P \rangle = (0.001\pm0.001)\langle d_{DC} \rangle + (0.580\pm0.009)}
\end{equation}
\begin{equation}
{\langle \varphi_{31} \rangle = -(0.008\pm0.005)\langle d_{DC} \rangle - (4.958\pm0.046)}
\end{equation}
their standard deviations being 0.232, 0.237 and 0.222, respectively. Obviously, the $\langle \log P \rangle$ and $\langle P \rangle$ are almost constant (the corresponding slopes being very small and zero within their $1\sigma$ error) on the average $\log P$ and $P$ of all our 454 RRab stars which are $-0.235\pm0.045$ (std) and $-0.585\pm0.059$ days (std), respectively, over against the LMC result mentioned above, since the LMC metallicity gradient is considered by FAW10 to be based on a $\log P$ gradient. On the contrary, $\langle \varphi_{31} \rangle$ shows a clear decrease with increasing distance from the SMC DC. A combination of the above relations with equations (2) and (3) of Paper I (A2 and A3) shows that the coefficients for the contribution terms of $d_{DC}$ from $P$ and $\varphi_{31}$ are 0.005 and 0.011, respectively, the latter being more than two times larger ($\langle [Fe/H]_{C09} \rangle  \sim -0.005 \langle d_{DC} \rangle ^{from}_{P} -0.011 \langle d_{DC} \rangle ^{from}_{\varphi_{31}}$). Thus, the metallicity gradient that was detected in our sample in the SMC seems to originate in a $\varphi_{31}$ gradient rather than in a $P$ (or $\log P$) gradient.}

\subsection{Conclusions}

A metallicity gradient of RR Lyrae stars is most likely bound to be correlated with radial gradients of their properties, such as $\log P$, $\varphi_{31}$, and/or their combinations. Despite the evidence of its existence, its detection is probably affected by selection effects that could either constitute its true origin or downgrade its substance, depending on the metal abundances and the actual distances from the SMC DC of non-included objects (undetected or excluded). A stronger confirmation could be provided by spectroscopically derived metal abundances of the large sample of RRab stars detected by OGLE-III in the SMC. Furthermore, according to Subramanian \& Subramaniam (2012, hereinafter SS12), who investigated the three-dimensional structure of the SMC using RR Lyrae and red clump stars, our present view of the SMC is like viewing only the central part of a sphere along the line-of-sight. This perspective, combined with the indication that the metal poor RRab stars have a larger scale height than the metal rich ones, as it was found in Paper I and thoroughly examined in Section 5.5 of the present work, implies that a robust examination of any metallicity gradient should await the spatially extended OGLE-IV survey, as well.


\section{Structural analysis of the SMC}

After a careful selection of our final sample of fundamental-mode RR Lyrae variables and the derivation of their metal abundances, we attempted a detailed structural analysis of the SMC. For this purpose, we derived their distance moduli, after determining their absolute magnitudes and applying corrections for interstellar extinction.

\subsection{Absolute magnitudes of the RRab stars}

We derived the absolute magnitudes ($M_V$) of the 454 RRab stars of our sample (which will be subsequently used for the estimation of the line-of-sight distances of these objects), using the method which was described in detail in Paper I (equations 10 and 11 there, i.e. A10 and A11, respectively). Equation (11) of Paper I, which was used for the transformation of the derived metallicities from the JK96 scale to the Harris (1996) scale (for determining $M_V$), is valid for metal abundances between -2.31 and -0.68 dex on the JK96 scale. Of our 454 stars, 21 have $[Fe/H]_{JK96}$ outside this range, although only 8 are still outliers after taking into account their $1\sigma$ error and are not expected to affect our results.

\par{The resulting absolute magnitudes for the individual objects are listed in Table 4 (Column 4), while in the last three lines of Table 5 we give some basic statistics for the absolute magnitudes, distance moduli and distances. The average value of the absolute magnitude for the 454 RRab stars is $M_V = 0.49\pm0.09$ mag (std), the minimum and maximum values being $0.26$ and $0.78$ mag, respectively.}

\subsection{Reddening of the SMC}

A correction for the interstellar extinction ($A_V$) is needed in order to derive the distance moduli of the RRab stars (equation 13 of Paper I, i.e. A13). The reddening values of Udalski et al. (1999, hereinafter U99), which were used in Paper I, have a spatial coverage limited to the central bar region of the SMC. The optical reddening map of Haschke, Grebel \& Duffau (2011, hereinafter HGD) is the most recent and suitable for our data as far as spatial coverage is concerned. These authors used the average colour of Red Clump (RC) stars on the colour-magnitude diagram to derive reddening values in $(V-I)$, i.e. $E(V-I)$, by adopting an average theoretical colour $(V-I)_0$ for the RC stars. The correction for the extinction ($A_V$) was derived using the equations $E(B-V)=E(V-I)/1.38$ and $A_V=3.32E(B-V)$ of Tammann, Sandage \& Reindl (2003) and Schlegel, Finkbeiner \& Davis (1998), respectively. A statistical error for the reddening of the individual stars was defined using the average $E(V-I)$ of fields of the HGD map within 10 arcmin from the position of the RR Lyrae stars.

\par{The average $E(V-I)$ for our sample was $0.041\pm0.013$ (std) with the minimum and maximum values being $0.013$ and $0.086$, respectively. These values correspond to an average $E(B-V)$ of $0.030\pm0.009$ (std) ranging from $0.010$ to $0.062$. Thus, the applied correction due to the interstellar extinction ($A_V$) was on average $0.099\pm0.031$ mag (std) and within the range from $0.032$ to $0.207$ mag. According to the reddening map of U99 for the central bar region of the SMC, the average $E(B-V)$ of the 11 fields of OGLE-II was $0.087\pm0.011$ (std), ranging between $0.070$ and $0.101$, the latter value being larger by $0.039$ than the corresponding one which has been statistically derived from the HGD map for the central bar (and the wing as well) of the SMC (i.e. 0.072). It has been found, however, that the U99 reddening for the LMC is overestimated by $0.028$ mag (Clementini et al. 2003, hereinafter C03; see also a brief discussion in subsection 6.1 of Paper I). Assuming that this is also the case for the SMC, the maximum values in $E(B-V)$ of U99 and HGD for the central regions of the SMC would be in good agreement.}

\par{It should be noted that a second source of error is present in the HGD reddening values, apart from the statistical one mentioned above, originating from the assumptions on which the determination of reddening is based, although it can only be qualitatively described. HGD adopted an average value of $0.89$ mag for the theoretical mean colour $(V-I)_0$ of RC stars, using the mean clump properties given by Girardi \& Salaris (2001, hereinafter GS01) and assuming an average metallicity of $z \sim 0.0025$ ($[Fe/H] \sim -0.9$ dex) for the SMC. However, such a simplified model introduces systematic errors, since the colour $(V-I)_0$ depends on the average age of the stellar populations and metallicity (GS01). The ages of the stellar populations of the SMC range from $\sim 0$ to $\sim 12$ Gyrs. Furthermore, an age gradient with decreasing average age towards the SMC center has been detected, e.g. by Gardiner \& Hatzidimitriou (1992). According to these authors, the average age ranges from 0.5 to 6.5 Gyrs close to the center and from 7.5 to 12.5 beyond $2.3$ kpc. According to Da Costa and Hatzidimitriou (1998) or, more recently, according to Kayser et al. (2009), there is an age-metallicity relation (AMR) for the SMC showing a decreasing average metal abundance with increasing average age. By combining the age gradient and the AMR (e.g. from Kayser et al.), the average metallicities ($[Fe/H]$) would range from -0.65 to -0.95 dex (the z ranging from 0.0045 to 0.0022) close to the center and from -1.05 to -1.55 (the z ranging from 0.0018 to 0.00056) at larger distances. Using the GS01 tables, these metallicity values would roughly result in an average $(V-I)_0$ ranging from $0.86\pm0.07$ mag (std) in the central region to $0.78\pm0.03$ mag (std) beyond $2.3$ kpc from the SMC center. Thus, the reddening of the SMC is probably underestimated by $0.02\pm0.05$ mag and $0.08\pm0.02$ mag (in $E(B-V)$) for its inner and outer regions, respectively, which, in turn, would imply systematic errors of $0.07\pm0.17$ and $0.27\pm0.07$, respectively, for the distance moduli of the SMC RR Lyrae stars. This rough calculation is only meant as an order of magnitude estimate of the (maximum) systematic error that can be introduced in our results due to the specific method used for the derivation of the reddenings by HGD.}

\subsection{Distance moduli of the RRab stars}

The individual distance moduli of the 454 RR Lyrae variables in our sample were derived using equation (13) of Paper I (i.e. A13). The resulting values and associated errors are shown in Table 4 (Column 6), while a histogram of the derived distance moduli is shown in Fig. 7. The average distance modulus for the SMC RR Lyraes (given in Table 5) is found to be $19.13\pm0.19$ (std), under the assumption that the distance modulus of the LMC is $18.52\pm0.06$ (see Paper I for a full discussion\footnotemark[3] \footnotetext[3]{Very recently, Storm et al. (2011) have found the LMC distance modulus to be $18.45\pm0.04$ using a sample of LMC Cepheids, while Ripepi et al. (2012) have reported an identical value of $18.46\pm0.03$ based on the $K_s$-band period - luminosity relations of the LMC Classical Cepheids. These estimates do not affect the average value which is adopted for the LMC distance modulus and is derived from other independent determinations listed in Paper I.}). The average of the errors (std) of the individual distance moduli, $\sigma_\mu$, is $0.08\pm0.02$ practically independent of the angular distance from the SMC DC. The systematic error due to the reddening uncertainties, which was described in Section 5.2, seems to be within the $\sigma_\mu$ of the majority of the stars which are lying within 2.3 deg from SMC DC (431 stars, i.e. $\sim 95 \%$, of the RRab stars of our sample), although it can be expected to affect the distance moduli of the few objects located in the outer regions of the SMC.

\par{In Paper I we found the average distance modulus of the SMC to be $18.90\pm0.18$ based on the RRab population of the central bar region and the U99 reddening maps. However (as also discussed in Paper I) the U99 reddenings seem to be systematically overestimated according to C03. Taking this into account renders the U99 reddening values compatible with the HGD ones and brings the average distance modulus of the Paper I RR Lyraes to $18.99\pm0.18$, in agreement with the present value (within the error). The distance modulus of a RR Lyrae star (derived with our method) depends on its mean apparent magnitude $A_0$ (i.e. the $m_V$), its metal abundance and the reddening. Our present sample contains RRab stars which are more metal poor by $\sim0.07$ dex on average than the Paper I sample. This would cause a small increase of the average distance modulus by $\sim0.02$ mag. On the other hand, the average $m_V$ (i.e. the average $A_0$) is identical in both samples. Thus, the differences in the reddening values adopted in the Papers I and II seem to be the cause of the systematic difference in the distance moduli; the average value adopted here (based on the HGD map) is lower by $\sim0.09$ in $E(B-V)$ than the corresponding average in the central bar from the U99 map, resulting in an increase in the average distance modulus by 0.23 mag.}

\par{Other recent independent determinations of the distance modulus of the SMC are listed in Paper I (Szewczyk et al. 2009; DS10; Kov\'{a}cs 2000; Harries, Hilditch \& Howarth 2003; Hilditch, Howarth \& Harries 2005; Crowl et al. 2001) and have an average of $18.91\pm0.11$ (std). North et al. (2010) have revised their previous estimate of $19.05\pm0.04$ (North, Gauderon \& Royer 2009) to a value identical to ours, i.e. $19.11\pm0.03$. Our estimate corresponds to a 0.61 mag difference between the SMC and LMC distance moduli, while the generally accepted range for the distance modulus difference lies between $\sim0.33 - 0.51$ mag (Matsunaga, Feast \& Soszy\'{n}ski 2011). Correcting for the systematic error in the reddening described in Section 5.2 would roughly lead to a reduction of the SMC distance modulus by at least $\sim0.08$ mag, bringing the difference from the LMC distance modulus to less than 0.53 mag.}

\par{The distance moduli of the RRab stars were used to derive their distances from us and, consequently, the average distance of the SMC, which was found to be $67\pm6$ kpc (std). These values are listed in Tables 4 and 5, respectively.}

\begin{figure}
\begin{center}
\leavevmode \epsfxsize=80mm
\epsffile{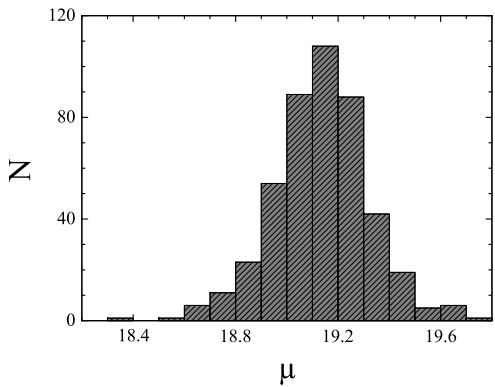}
\end{center}
\caption{Distribution of the individual distance moduli of 454 RRab.}
\end{figure}

\subsection{The line-of-sight depth of the SMC and its variations}

The line-of-sight (LOS) depth of the SMC has been the subject of numerous investigations. The current sample of fundamental-mode RR Lyrae stars is spatially distributed over a large area, contrary to the corresponding sample of Paper I which was restricted to the innermost regions of the SMC, being thus suitable to investigate the depth of the SMC and possible variations. Following Paper I but improving the method described therein, i.e. by taking into account the uncertainties due to extinction, we used the standard deviation of the average of the distances (derived from the distance moduli) of the individual RRab stars, $\sigma_{obs}$, where $\sigma_{obs}^2 = \sigma_{int}^2 + \sigma_{err}^2 + \sigma_{ext}^2$. The second additive term $\sigma_{err}$ is the average value of the standard deviations of the individual distances (given in the last column of Table 4), while $\sigma_{ext}$ is the average value of the standard deviations of the individual corrections for the extinction ($A_V$). Thus, $\sigma_{int}$ is taken to be "intrinsic", that is, due to the LOS depth of the sample and given by $\sigma_{int} = \sqrt{\sigma_{obs}^2 - \sigma_{err}^2 - \sigma_{ext}^2}$, which yields a $\pm 1\sigma$ LOS depth of $\sigma_{int}=5.3\pm0.4$ kpc (std). Other independent determinations are listed in Paper I.

\par{Our current analysis provides an independent confirmation of a $1\sigma$ LOS depth of $\simeq 5$ kpc, apparently shared by old and intermediate-age populations in the SMC. It would be more interesting, though, to investigate possible spatial variations of the SMC LOS depth, as it was done by SS09 who used a large sample of RC stars. The extended size of the sample of RRab stars and the larger spatial coverage across the face of the SMC, contrary to the dataset of Paper I, allows us to proceed to such a structural analysis and examination of the LOS depth variability.}

\par{The SMC has been proposed to be deeper in the NE region by HH89 and Gardiner \& Hawkins (1991), although they referred to a region beyond the spatial limits of the OGLE-III survey. To investigate this possibility in the area covered by OGLE-III, we examined three different samples of RRab stars, lying in three regions (north-eastern, NE; central; south-western, SW) as shown in Fig. 4. The $1\sigma$ LOS depth in the NE region was found to be $5.36\pm0.21$ kpc, while the corresponding value for the SW region was found to be $4.32\pm0.19$ kpc. Interestingly, the LOS depth in the central region was found to be $5.65\pm0.45$ kpc, suggesting thus the existence of a thicker structure. We also examined possible radial variations of the LOS depth using the four regions mentioned in Section 4 and shown in Fig. 4. The values corresponding to the central region, the inner ring, the outer ring and the outer region were found to be $6.38\pm0.60$ kpc, $5.55\pm0.12$ kpc, $5.23\pm0.14$ kpc and $4.46\pm0.24$ kpc (Table 7), showing a clear decrease with increasing angular distance from the SMC DC. Recently, SS12 presented similar results using RC stars. They found an average $1\sigma$ LOS depth of $4.57\pm1.03$ kpc for the SMC, a prominent feature of larger depth in the central region and an increased depth towards the NE region. The existence of a possible central substructure, as also suggested by SS09, and the distribution of populations of different metal abundances are examined in the next subsection.}

\subsection{Metallicity and structure}

Given the size of our sample of RRab stars, we can combine LOS distances and metal abundances, which allows us to investigate the possible presence of different structures consisting of different populations in the SMC, following the discussion in Paper I. We divided the sample of 454 RRab stars into 7 subsamples with increasing average metal abundance. The first group consisted of stars with $[Fe/H]_{C09}<-2.25$ dex, the next five were limited by metallicities from -2.25 to -1.00 dex with a step of 0.25 dex (on the C09 scale), while the last group included all stars with $[Fe/H]_{C09}>-1.00$ dex. The $1\sigma$ LOS depths of these groups are plotted against the corresponding average metallicities in Fig. 8. This figure is suggestive of different structures corresponding to different metal abundances on average. The LOS depths of the two extreme populations (metal poor and metal rich stars) in the four elliptical regions mentioned in Section 4 and listed in Table 7 corroborate this result; the metal rich stars show small variations of their LOS depth around $\sim 5$ kpc with a small increase in the inner region while the metal poor stars have a larger increase in LOS depth towards the SMC center.

\begin{figure}
\begin{center}
\leavevmode \epsfxsize=80mm
\epsffile{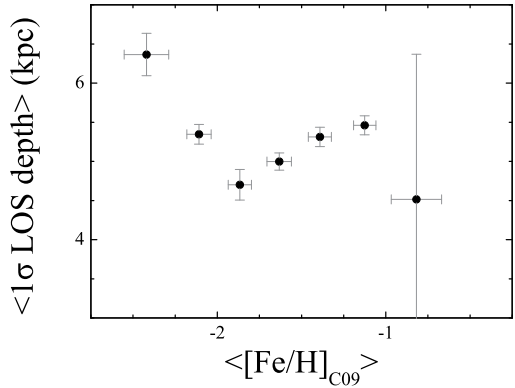}
\end{center}
\caption{Average $1\sigma$ LOS depth versus average metal abundance (on the C09 scale) for 7 subgroups of RRab stars. The metal poor and metal rich stars seem to constitute different structures.}
\end{figure}

\par{In order to further investigate this suggested trend, we used 25 elliptical regions, based on the isopleths of G09 and centered on the SMC DC, with increasing major axis length, starting from 0.7 to 3.0 deg with a step of 0.1, while the $25^{th}$ region was extended beyond 3.0 deg. The $1\sigma$ LOS depths for these areas were derived for all stars within the same area and also for the corresponding extreme subsamples of metal poor ($[Fe/H]_{C09}<-2.0$ dex) and metal rich ($[Fe/H]_{C09}>-1.4$ dex) RRab stars. Our results are illustrated in Fig. 9 with thick black, thin grey and thin black lines, respectively. Some interesting features are revealed.
\begin{enumerate}
\item {} The metal rich objects (thin black line) seem to occupy a much narrower structure, possibly a thick disk, with an average $1\sigma$ LOS depth slightly greater than 5 kpc beyond $\sim 1.5$ deg. In the innermost region ($<1.5$ deg) there seems to be a much thicker structure reminiscent of a bulge with a characteristic radius which may be estimated by an exponential fitting. The corresponding mathematical formula is described below.
\begin{equation}
{d_{los} = (45\pm14)e^{-\frac{a}{(0.193\pm0.016)}} + (5.20\pm0.01)}
\end{equation}
where $d_{los}$ (kpc) is the $1\sigma$ LOS depth and $a$ (deg) is the semi major axis of the ellipse. The width ($2\sigma$ depth) of the thick disk would then be $2\times(5.20\pm0.01)=10.40\pm0.02$ kpc. A rough estimation of the size (radius) of a possible bulge would result (using the equation above) from the angular distance ($a$) where the LOS depth falls to the disk limit value (within its error), i.e. $a = 1.62\pm0.24$ deg. This semi major axis length corresponds to a maximum distance of $1.78\pm0.26$ deg from the DC, after restoring the geometry of the RA-Dec map, or $2.09\pm0.81$ kpc, assuming a symmetrical distribution and a distance scale where the SMC is on average $67\pm6$ kpc away from us.
\item {} The metal poor objects on the other hand occupy a much thicker structure probably deeper than 16 kpc ($2\sigma$ depth) near the center. This could be interpreted as a spheroidal structure such as a halo. Its depth along the line of sight varies between $\sim 16$ kpc (or more) to $\sim 12$ kpc with increasing distance from the SMC DC. It should be noted that the inverse slope of the metal poor line in the innermost ellipses indicates a lack of metal poor objects in these regions of the SMC (below 1 deg). This is expected since these stars have smaller amplitude light curves and greater $D_m$, being thus more difficult to be detected in the innermost regions of the SMC (as discussed in Section 4).
\item {} An exponential fitting to the depth variation of the overall population (thick black line in Fig. 9) would result in a maximum $1\sigma$ LOS depth of $9.24\pm0.34$ kpc in the SMC DC, significantly larger than the nearly constant value of $5.30$ kpc in the outer regions. This is consistent with the results of SS12 from RC stars. Furthermore, following SS12, we attempted to derive the axes ratios of an ellipsoid that would fit our data. We used RRab stars within spherical cells of different radii, i.e. 2.5, 3.0 and 3.5 kpc ($\sim$ 2.1, 2.6 and 3.0 deg, respectively). The corresponding ratios were 1:1.21:1.57, 1:1.18:1.53 and 1:1.23:1.80 (the longest axis being along the line of sight and the number of stars being 118, 161 and 200, respectively). As also mentioned by SS12, who found similar results (table 5 in SS12), the estimated ratios strongly depend on the data coverage.
\end{enumerate}
}

\par{A robust description of the bulge and halo characteristics must await the OGLE-IV observations of the full extent of the SMC on the RA-Dec plane. In that sense, the bulge size estimated above perhaps should be considered as a lower limit. Furthermore, it would be very important to combine this type of data with kinematical information (radial velocities) of RR Lyrae stars to investigate if these structures are indeed consistent kinematically with a halo, disk and bulge components.}

\begin{figure}
\begin{center}
\leavevmode \epsfxsize=80mm
\epsffile{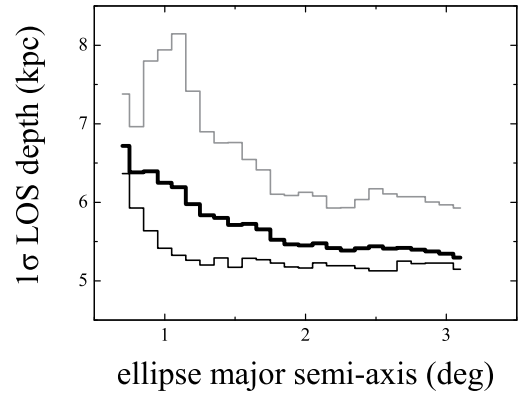}
\end{center}
\caption{Average $1\sigma$ LOS depth of 25 groups of RRab stars belonging to elliptical areas with increasing major axis, based on the isopleths of Gonidakis et al. and centered on the SMC dynamical center. The thin black, thin grey and thick black lines correspond to metal rich, metal poor and all RRab stars.}
\end{figure}


\section{Summary \& Conclusions}

This is the second of a series of papers presenting metal abundances of RR Lyrae variable stars in the SMC. We have performed Fourier decomposition analysis of 8- and 13-year V-band light curves of a carefully selected sample of 454 fundamental-mode RR Lyrae variables, detected by OGLE in the SMC and listed in the OIII-CVS. Their Fourier decomposition parameters were used to derive metal abundances and distance moduli, as well as to perform a chemical and structural analysis of the SMC, which was hampered in Paper I due to the limited size and extent of the available sample in the central region of the SMC.

\par{The average metal abundance of these RRab stars on the new scale of C09 is found to be $\langle [Fe/H]_{C09} \rangle=-1.69\pm0.41$ dex (std, with a standard error of $0.02$ dex). Furthermore, the extended region of the SMC is populated by more metal poor objects, compared to the central bar region (Paper I). The distribution of our RRab stars of different metal abundances on the Bailey Diagram showed a clear anti-correlation between amplitude and period, as predicted by theoretical models, as well as the expected (from the models) displacement between the loci of low and high metallicity RRab stars, and the flattened extension of the curve for higher metallicity objects. The bulk of the RR Lyrae variables with intermediate metallicities possibly constitute an intermediate Oosterhoff population.}

\par{A tentative metallicity gradient was detected, with increasing average metal abundance towards the SMC DC, originating from a relative surplus of high metallicity objects in the inner regions and mainly related to a radial gradient of the Fourier parameter $\varphi_{31}$. The average metal abundances of the RRab stars show a slope of $-0.013\pm0.007$ dex/kpc relative to their average distances from the SMC DC. Similar results were found by FAW10 for the LMC RRab variables. Selection effects were examined through the ratio $N_{RRab}/N_{RRc}$ of the populations of RRab and RRc stars in the same region and other aspects were also discussed. Although they may not play a crucial role, their importance is strongly dependent on the metal abundances and the actual distances from the SMC DC of any non-included objects (either undetected by OGLE or excluded by selection criteria based on the quality of their light curves). Spectroscopically derived metallicities of a large, spatially extended, sample of RR Lyrae stars in the SMC are needed for a robust confirmation of the existence of the metallicity gradient and to clarify its origin.}

\par{The distance modulus of the SMC was found to be $\langle \mu \rangle=19.13\pm0.19$ (std), in a distance scale where the distance modulus of the LMC is $\mu_{LMC} = 18.52\pm0.06$. The distances to individual RRab stars were used to study the LOS depth in the SMC and its variations. The SMC was found to have an average LOS depth of $5.3\pm0.4$ kpc (std), also being deeper in the north-eastern region (compared to the south-western one) by $1.04\pm0.41$ kpc. Moreover, there is a clear indication of a thicker structure in the inner regions of the SMC, reminiscent of a bulge. Metal rich and metal poor objects in the sample seem to belong to different dynamical structures. The former have smaller scale height and may belong to a thick disk, its width ($2\sigma$ depth) being $10.40\pm0.02$ kpc, and a bulge whose size (radius) is estimated to be $2.09\pm0.81$ kpc. The metal poor objects seem to belong to a halo whose ($2\sigma$) depth along the line of sight extends over 16 kpc in the inner regions of the SMC, while $\sim 12$ kpc is a rough estimation for the outer regions. Combination with kinematics of RR Lyrae stars are needed to clarify these issues.}


\section*{ACKNOWLEDGMENTS}

We would like to thank the anonymous referees of Papers I and II, whose corrections and comments contributed substantially to the improvement of both papers. We are also grateful to OGLE for having their data publically available. E. Kapakos thanks R. Haschke for providing the HGDJ's metallicities of the individual RR Lyrae stars of the SMC that facilitated a thorough and useful comparison between the two methods.



\section*{SUPPORTING INFORMATION}

Additional Supporting Information may be found in the online version of this article:\\
\\
{\bf Table 1.} Fourier decomposition parameters for 1887 RRab stars derived from data of the OGLE phases II \& III in the V-band.\\
{\bf Table 2.} Fourier decomposition parameters for 169 RRc stars derived from data of the OGLE phases II \& III in the V-band.\\
{\bf Table 4.} Metal abundances, absolute magnitudes, distance moduli and distances for the 454 RRab stars.\\


\appendix

\section{Mathematical formalism}

The mathematical formalism, following the analysis and the techniques presented in Paper I, is summarized here with a brief description.

\par{The light curves in the V-band of the RR Lyrae variables are fitted using Fourier series of sine functions (eq. 1 of Paper I):
\begin{equation}
{mag=A_0+\Sigma_{j=1}^{4} A_j
\sin\left(j\frac{2\pi}{P}t+\varphi_{j}\right)}
\end{equation}
where $P$ is the period, $t$ the the time of the observation, $A_0$ is the mean apparent magnitude of each star, while $A_j$ and $\phi_j$ are the amplitudes and phases (reduced into the normal interval $[0,2\pi]$) for $j=1,2,3,4$.}

\par{The metal abundances of the RRab stars in the JK96 and C09 scales are derived from the important Fourier parameter $\phi_{31}$ using the equations below (eq. 2 and 3 of Paper I):
\begin{equation}
{[Fe/H]_{JK96}=-5.038-5.394 P+1.345\varphi_{31}}
\end{equation}
\begin{equation}
{[Fe/H]_{C09}=(1.001 \pm 0.050)[Fe/H]_{JK96}-(0.112\pm0.077)}
\end{equation}
The corresponding standard deviations are (eq. 4 and 5 of Paper I):
\begin{eqnarray}
\sigma_{[Fe/H]_{JK96}}^2  =
1.809\sigma_{\varphi{31}}^2+2K_{12}P+2K_{13}\varphi_{31}\nonumber\\
+ 2K_{23}P\varphi_{31}+K_{11}+K_{22}P^2+K_{33}\varphi_{31}^2
\end{eqnarray}
where the period error has been omitted and\\
$K_{11}=0.08910$ ~~ $K_{22}=0.02529$ ~~ $K_{33}=0.00374$ \\
$K_{12}=0.00116$ ~~ $K_{13}=-0.01753$ ~~ $K_{23}=-0.00289$\\
\begin{equation}
{\sigma_{[Fe/H]_{C09}}^2=0.002[Fe/H]_{JK96}^2+1.002\sigma_{[Fe/H]_{JK96}}^2+0.006}
\end{equation}
}

\par{The metal abundances of the RRc stars in the CG and C09 scales are derived from the important Fourier parameter $\phi_{31}$ using the equations below (eq. 6 and 7 of Paper I):
\begin{equation}
{[Fe/H]_{CG}=0.0348\varphi_{31}^2+0.196\varphi_{31}-8.507P+0.367}
\end{equation}
\begin{equation}
{[Fe/H]_{C09}=(1.141 \pm 0.042)[Fe/H]_{CG}+(0.003\pm0.061)}
\end{equation}
The corresponding standard deviations are (eq. 8 and 9 of Paper I):
\begin{equation}
{\sigma_{[Fe/H]_{CG}}^2=(0.0696\varphi_{31}+0.196)^2\sigma_{\varphi_{31}}^2 +72.369\sigma_{P}^2}
\end{equation}
\begin{equation}
{\sigma_{[Fe/H]_{C09}}^2=0.002[Fe/H]_{CG}^2+1.141\sigma_{[Fe/H]_{CG}}^2+0.004}
\end{equation}
}

\par{The absolute magnitudes of the RR Lyrae stars are derived through the above linear relation (eq. 10 of Paper I):
\begin{eqnarray}
M_V = (0.214\pm0.047)([Fe/H]_{H}+1.5)\nonumber\\
+ (19.064\pm0.017) - \mu_{LMC}
\end{eqnarray}
where [Fe/H] is on the Harris (1996) scale and $\mu_{LMC}$ is the LMC distance modulus.
These metalicities are calculated from the corresponding ones in the JK96 and the CG scales for the RRab and RRc stars, respectively, using the following transformations (eq. 11 and 12 of Paper I):
\begin{equation}
{[Fe/H]_{H}=(0.992 \pm 0.043)[Fe/H]_{JK96}-(0.175\pm0.068)}
\end{equation}
\begin{equation}
{[Fe/H]_{H}=(1.064 \pm 0.042)[Fe/H]_{CG}-(0.048\pm0.061)}
\end{equation}
}

\par{The distance moduli of the RR Lyrae are derived from their absolute and apparent magnitudes ($M_V$ and $m_V$, i.e. $A_0$, respectively), by taking into account the correction for the interstellar extinction, as well (eq. 13 of Paper I):
\begin{equation}
{\mu = m_V - M_V - A_V}
\end{equation}
}

\par{Extended discussions on all the above equations are available in Paper I for interested readers.}

\end{document}